\documentclass[aps,amssymb,amsmath,prd,twocolumn,
showpacs,preprintnumbers,superscriptaddress,nofootinbib,floatfix]{revtex4-1}
\usepackage{graphicx}
\usepackage{orcidlink} 
\usepackage{color}
\usepackage{times}
\usepackage{amsmath}
\usepackage[normalem]{ulem}
\usepackage{inputenc}
\usepackage{bm}
\usepackage{multirow}
\usepackage{float}
\usepackage{url}
\usepackage{natbib}
\usepackage{wrapfig}

\hypersetup{colorlinks=true,linkcolor=red,urlcolor=blue,citecolor=blue}

\newcommand{\Msun}{\,{M}_{\odot}}
\def\fm3{\;\text{fm}^{-3}}

\newcommand{\ie}{i.e.,~}

\begin{document}

\title{Constraining the Equation of State of Neutron Stars with third-generation Gravitational Wave detectors}

\author{Zhenyu Zhu \orcidlink{0000-0001-9189-860X}}
\author{Richard O'Shaughnessy \orcidlink{0000-0001-5832-8517}}
\affiliation{Center for Computational Relativity and Gravitation, Rochester Institute of Technology, Rochester, NY 14623, USA;\\}

\begin{abstract}
    We investigated the impact of the number of binary neutron star merger events and neutron star (NS) mass distributions on constraining the equation of state (EoS), tidal deformability and radius of NSs, as well as the nuclear parameters, using binary neutron star inspiral gravitational wave signals with third-generation detectors. We generate simulated gravitational wave signals and compute the Fisher information matrix for each event after the number of events and mass distribution models are given. The covariance of EoS parameters is obtained by holding the non-EoS waveform parameters fixed at their injected values and accumulating the Fisher matrix over all events. Finally, the posterior samples of EoS, tidal deformability, radius and nuclear parameters are generated based on this covariance matrix. We find that larger number of events lead to tighter constraints due to more observed events and data accumulation, as expected given the increased number of detections. Meanwhile, we note that the mass distribution plays a more important role in constraining the EoS. We compare a realistic bimodal Gaussian distribution, a uniform distribution and another uniform including sub-solar mass NSs. The results show that the uniform distribution yields tighter constraints than bimodal models because it includes more low-mass and massive NSs. This also implies that events near $1.4\,\Msun$ provide partly redundant information about the EoS. Additionally, including sub-solar mass NSs can further improve the constraints by significantly reducing the EoS uncertainties at sub-saturation densities and inner-crust, highlighting the importance of sub-saturation density EoS.
\end{abstract}

\maketitle

\section{Introduction}

The equation of state (EoS) of dense matter is a crucial ingredient for understanding both the strong interactions and the properties of neutron stars (NSs). It specifies the pressure of dense matter as a function of density, temperature, and composition, determined by the underlying nuclear interactions. However, the EoS at supranuclear densities remains poorly constrained due to the challenges in both theoretical modeling and experimental measurements. Nowadays, with advancements in astrophysical observations including the gravitational wave (GW) detections~\citep{2017PhRvL.119p1101A, 2019PhRvX...9a1001A, 2018PhRvL.121p1101A, 2020ApJ...892L...3A}, X-ray measurements~\citep{2019ApJ...887L..24M, 2019ApJ...887L..21R, 2021ApJ...918L..28M, 2021ApJ...918L..27R, 2024ApJ...971L..20C}, and radio pulsar timing~\citep{2010Natur.467.1081D, 2020NatAs...4...72C, 2013Sci...340..448A, 2016ApJ...832..167F}, accurate constraints on the EoS are expected in the near future. From an astrophysical perspective, the EoS serves as the final component in hydrodynamical equations, which are essential for describing the macroscopic behavior of NSs. Hence, it impacts fundamental properties such as mass, radius, moment of inertia, and tidal deformability, and plays a key role in the dynamics of binary mergers and core-collapse supernovae. Constraining the dense matter EoS is therefore of great importance and has been a long-standing goal in both nuclear physics and astrophysics.

In most of astrophysical scenarios, where static or quasi-static conditions apply, the neutron star matter resides in a cold and $\beta$-equilibrium state. Under such conditions, the EoS can be effectively described by a barotropic relation, i.e., pressure depends solely on the density. Various parameterizations of these one-dimensional EoS have been proposed, including piecewise polytropes~\citep{2018PhRvL.120q2703A, 2018PhRvL.120z1103M, 2019ApJ...881...73W, 2025arXiv251205315S}, spectral decomposition~\citep{2010PhRvD..82j3011L, 2022PhRvD.105f3031L, 2022ApJ...926..196H, 2024PhRvD.110h3030L, 2024arXiv241014674Y, 2024arXiv240715753V}, Gaussian process~\citep{2018PhRvL.121p1101A, 2024PhRvD.109b3020L, 2025PhRvD.112f3003L, 2025CQGra..42t5008N, 2025PhRvD.112j3023F}, and speed-of-sound models~\citep{2021ApJ...919...11H, 2022ApJ...939L..34A, 2023ApJ...950...77H}. However, these barotropic parameterizations generally lack direct connections to nuclear properties, making it difficult to incorporate complementary information from nuclear experiments and theoretical nuclear calculations. Therefore, these nuclear physics informed models or parameterizations, such as relativistic mean-field (RMF) theory~\citep{2018PhRvC..97c5805Z, 2018ApJ...862...98Z, 2023PhRvC.108b5809Z, 2023ApJ...943..163Z, 2020ApJ...897..165T,2023ApJ...943..163Z, 2025PhRvD.112f3003L}, quark mean-field (QMF) model~\citep{2018PhRvC..97c5805Z, 2018ApJ...862...98Z, 2019PhRvC..99b5804Z, 2023PhRvC.108b5809Z, 2025PhRvD.112d3018T}, Skyrme-Hartree-Fock (SHF) theory~\citep{2018PhRvC..98e4618W, 2024PhRvC.109e4623W, 2025PhRvC.111e4605W}, and Taylor expansions of nuclear matter and symmetry energy~\citep{2018ApJ...859...90Z, 2021ApJ...921..111Z, 2024PhRvD.110j3040L, 2025arXiv250415893W}, have been widely employed for more comprehensive description of dense matter. These models not only provide flexible EoS parameterizations but also establish links to nuclear properties, enabling their direct inference from observational and experimental data.

The EoS of neutron star matter has a one-to-one correspondence with the mass-radius (M-R) relation, which can be obtained by solving the TOV equations. Therefore, accurate and simultaneous measurements of NS masses and radii are highly informative for the EoS. X-ray pulsar observations enabled such measurements and have significantly advanced our ability to constrain the EoS~\citep{2019ApJ...887L..24M, 2019ApJ...887L..21R, 2021ApJ...918L..28M, 2021ApJ...918L..27R, 2024ApJ...971L..20C}. In addition, the measurements of massive neutron stars provide valuable constraints by setting a lower bound on the maximum mass~\citep{2010Natur.467.1081D, 2020NatAs...4...72C, 2013Sci...340..448A, 2016ApJ...832..167F}. The moment of inertia (MOI) of NSs, which depends on the EoS, can be measured through the pulsar timing and provide complementary information~\citep{2020MNRAS.497.3118H, 2021ApJ...915L..12F, 2021PhRvX..11d1050K}. Kilonova emissions powered by the radioactive decay of $r$-process elements in the ejecta from binary neutron star (BNS) mergers, can likewise correlate with the EoS~\citep{2017ApJ...850L..19M, 2018ApJ...852L..29R, 2019MNRAS.489L..91C, 2021MNRAS.505.1661B, 2022ApJ...926..196H, 2023ApJ...943..163Z}. The stiffness of EoS affects the radius and tidal deformability of NSs, which in turn influence the violence of collision during the merger and impact the ejecta properties such as mass, velocity, and composition. Thus, kilonova observations may also offer valuable hints about the nature of dense matter.

Meanwhile, the progress in nuclear experiments and theoretical calculations has yielded further additional insights into the properties of dense matter. For instance, the measurements of neutron-skin thickness of nuclei~\citep{2021PhRvL.126q2502A, 2022PhRvL.129d2501A} and data from heavy-ion collisions~\citep{2002Sci...298.1592D, 2006PrPNP..56....1F, 2009PhRvL.102l2701T, 2014NuPhA.922....1D, 2016PhRvC..94c4608R_etal, 2021PhRvC.103a4616L, 2021PhRvC.103a4616L} have placed important constraints on the symmetry energy and pressure. Recent advancements in the chiral effective field theory have also provided reliable constraints of the EoS at low densities, which can be used as a benchmark for other EoS models~\citep{2016PhRvC..93e4314D, 2024A&A...687A..44D, 2024PhRvD.109j3015S, 2025arXiv250516929K}. The combination of more complementary observations and experiments is expected to yield a more comprehensive understanding of dense matter.

Gravitational wave (GW) detection from binary neutron star mergers is another powerful tools for constraining the
neutron star properties, especially given the anticipated measurement accuracy enabled by  third-generation (3G) GW
detectors, \ie Einstein Telescope (ET) and Cosmic Explorer (CE)~\citep{2010CQGra..27h4007P, 2012CQGra..29l4013S,
  2020JCAP...03..050M, 2019BAAS...51g..35R}. The first detection of the BNS merger event
GW170817~\citep{2017PhRvL.119p1101A, 2019PhRvX...9a1001A} by LIGO/Virgo detectors~\citep{2015CQGra..32g4001L, 2015CQGra..32b4001A} have provided the measurement of tidal deformability (TD),
which significantly improved the EoS and NS radii constraints~\citep{2018PhRvL.121p1101A, 2018ApJ...857L..23R,
  2018PhRvL.120q2703A, 2018PhRvD..98f3020Z, 2018PhRvL.120z1103M, 2019ApJ...881...73W}. 
Later, another BNS event, GW190425~\citep{2020ApJ...892L...3A}, was detected, although no further EoS information was obtained from this event. 
The tidal deformability, which is uniquely determined by the EoS, is encoded in the inspiral phase of the GW signal through its impact on the orbital
dynamics~\citep{2019PhRvD.100d4003D}. Furthermore, the upcoming 3G detectors are expected to detect a large amount of
BNS mergers with sufficiently high signal-to-noise ratio (SNR) to resolve the effects of tidal deformability. By
accumulating the information from numerous BNS merger events, these future observations would provide tight constraints on the mass-tidal deformability relation, thereby enabling accurate inference of the EoS and nuclear properties of dense matter~\citep{2022arXiv220501182G, 2022PhRvD.106l3529G, 2024PhRvD.110d3013W}.

The mass distribution of neutron stars is an important factor in EoS
inference~\citep{2020arXiv200101747W,2024arXiv241014674Y,2025arXiv251212130A}. Observational data generally suggest
that, averaging over different channels which produce detectable neutron stars, 
the net mass distribution is bimodal, with a sharp cutoff representing the maximum neutron star
mass~\citep{2018MNRAS.478.1377A, 2025arXiv251212130A}. This cutoff is EoS-dependent, which highlights the significance
of pulsar mass measurements for constraining the EoS~\citep{2024PhRvD.109d3052F}. Moreover, EoS inference from the NICER
mass-radius measurements necessarily  relies on assumptions about the mass distribution, which can be degenerate with EoS parameters. Inferring
the mass and EoS independently may introduce biases into posterior conclusions about nuclear matter~\citep{2020arXiv200101747W,2025PhRvD.112b3045B}. Finally, given the
large number of detectable event from 3G detectors, the mass distribution will also play an important role in
distinguishing neutron stars from low-mass black holes. 

Observational data provides stringent constraints on any model for the EoS.   Ideally,  these observational inputs can
be combined hierarchically to construct a Bayesian EoS posterior, using inputs obtained for each observation  using  for
example the Markov Chain Monte Carlo (MCMC)
method~\citep{2015PhRvD..91d2003V, 2020MNRAS.499.3295R, 2023PhRvD.107b4040W} or equivalent quadrature
techniques~\citep{2018arXiv180510457L}.  These detailed analyses of each observation should enable the most reliable and
accurate EoS posterior.  However, in these scenarios of forecasting parameter uncertainties from large number
of simulated data, such as the BNS GW signals expected to be detected by 3G detectors, both constructing and employing
detailed inferences from the many detected sources becomes computationally expensive and impractical.
Particularly for forecasting the expected performance of future instruments, an alternative and efficient approach is
the Fisher information matrix (FIM) method, which provides the covariance matrix of the Bayesian posterior probability
distribution of parameters under the linear
approximation~\citep{1994PhRvD..49.2658C,1995PhRvD..52..848P,2013PhRvD..87b4004C, 2014PhRvD..89f4048O,2020ApJS..250....6W, 2022ApJS..263....2I, 2022ApJ...941..208I, 2024PhRvD.109j3035H}.
For loud sources, the FIM provides an increasingly accurate estimate of the posterior derived through full Bayesian
methods, while remaining computationally tractable at scale.
Considering the sensitivity of 3G GW detectors, tens to hundreds of BNS merger events are expected to be detected per day with sufficiently high SNR. The accumulation of the observations can lead to very small uncertainties in EoS parameters. Therefore, the FIM method offers both sufficient accuracy and computational efficiency in this scenario.

This paper is organized as follows. First, in Section ~\ref{sec:methodology}, we describe the methodology used in this
work, including the formulations of relativistic mean field (RMF) theory, the crustal EoS parameterization, the
variational formulation of TOV and tidal equations, the details of number of events and mass distribution models, and the
fisher information matrix (FIM).  In, Sec.~\ref{sec:results}, the posteriors of EoS and NS properties are presented and compared under different number of events and mass distribution models, their implications are discussed in details. Finally, we summarize and conclude in Sec.~\ref{sec:conclusion}.

\section{Methodology}
\label{sec:methodology}

\subsection{Relativistic mean field theory}
Relativistic mean field theory is a theoretical framework used to describe the properties of nuclear matter. In this approach, mesons mediate the nuclear interaction through their coupling with nucleons. The theory begins with a Lagrangian that defines these interactions~\citep{2018PhRvC..97c5805Z,2018ApJ...862...98Z, 2021Univ....7..399A,2022PhRvC.106e5804A, 2023PhRvC.108b5809Z,2023ApJ...943..163Z}
\begin{eqnarray}
  \label{eq:Lagrangian}
    \mathcal{L}& = & \overline{\psi}\left(i\gamma_\mu \partial^\mu - M_N + g_\sigma \sigma - g_{\omega}\omega_\mu \gamma^\mu - g_{\rho}\bm{\rho}_\mu\cdot \bm{\tau}\gamma^\mu\right)\psi  \nonumber \\
           && \overline{\psi}_{e}(i\gamma_\mu \partial^\mu - m_e)\psi_e + \frac{1}{2}\partial_\mu\sigma \partial^\mu\sigma - \frac{1}{2}m_\sigma^2 \sigma^2 \nonumber \\
           && - \frac{1}{3} g_2\sigma^3 - \frac{1}{4}g_3\sigma^4 - \frac{1}{4}\omega_{\mu\nu}\omega^{\mu\nu} + \frac{1}{2}m_\omega^2\omega^2 + \frac{1}{4}c_3 \omega^4\nonumber \\
           & & - \frac{1}{4}\bm{B}_{\mu\nu}\cdot \bm{B}^{\mu\nu} + \frac{1}{2}m_\rho^2\bm{\rho}^2\ + \frac{1}{2}g_{\omega}^2\omega^2 \Lambda_v g_{\rho}^2\bm{\rho}^2 \; ,
\end{eqnarray}
where
\begin{eqnarray}
  \label{eq:omega_rho}
  \omega_{\mu\nu} & = & \partial_\mu \omega_\nu - \partial_\nu \omega_\mu\, ,\ \ \ \omega^2=\omega_\mu \omega^\mu \\
  \bm{B}_{\mu\nu} & = & \partial_\mu \bm{\rho}_\nu - \partial_\nu \bm{\rho}_\mu\, ,\ \ \ \bm{\rho}^2= \bm{\rho}_\mu \cdot \bm{\rho}^\mu \; .
\end{eqnarray}
The $\psi$, $\psi_e$ $\sigma$, $\omega_\mu$, and $\bm{\rho}_\mu$ denote the field of nucleon, electron, the $\sigma$, $\omega$, and $\rho$ mesons, respectively. The masses of mesons are represented by $m_\sigma$, $m_\omega$ and $m_\rho$. This model contains 7 coupling constants, $g_\sigma$, $g_\omega$, $g_\rho$, $g_2$, $g_3$, $c_3$, and $\Lambda_v$, which have no direct physical interpretation in terms of measurable nuclear observables. Therefore, we convert these coupling constants into a set of effective parameters of nuclear matter that can be constrained by experimental data, \ie the nuclear properties at saturation density. For the purpose of both simplicity and convenience, we choose the following 7 effective parameters: the saturation density $n_0$, the binding energy $E_0$, incompressibility $K_0$, skewness $J_0$, symmetry energy $E_{\rm sym}$, symmetry energy slope $L_{\rm sym}$, and the Dirac effective mass $M^*_N$. 

We recall the definition of symmetry energy $E_{\rm sym}(n_{_B})$ and these nuclear properties:
\begin{eqnarray}
  \label{eq:esym_def}
  E_{\rm sym}(n_{_B}) & = & \frac{1}{8}\frac{\partial^2 E(n_{_B}, Y_p)}{\partial Y_p^2} \bigg|_{Y_p=1/2}, \\
  E(n_{_B}, 1/2) & = & E_0 + \frac{K_0}{2} \left( \frac{n_{_B}-n_0}{3n_0} \right)^2 \nonumber \\
      & & + \frac{J_0}{6} \left( \frac{n_{_B}-n_0}{3n_0} \right)^3  + O\biggl((n_{_B}-n_0)^4\biggr), \\
  E_{\rm sym}(n_{_B},0 ) & = & E_{\rm sym} + L_{\rm sym} \left( \frac{n_{_B}-n_0}{3n_0} \right) \nonumber \\
      & & + \frac{K_{\rm sym}}{2} \left( \frac{n_{_B}-n_0}{3n_0} \right)^2  + O\biggl((n_{_B}-n_0)^3\biggr). \nonumber \\
\end{eqnarray}
where $E(n_{_B}, Y_p)$ represents the energy per baryon of nuclear matter. Among these parameters, $n_0$, $E_0$, and
$M^*_N$ can be directly used to determine the Fermi momentum of nuclear matter, which would significantly simplify the
root-finding process. The expressions for the remaining nuclear property parameters can be written in terms of the
coupling constants as (note that there is a typo in the expression for $L_{\rm sym}$ in Eq. (10) of Ref.~\citep{2023ApJ...943..163Z}):
\begin{eqnarray}
  \label{eq:esym}
  E_{\rm sym} & = & \frac{p_{\rm F}^2}{6E_{\rm F}} + \frac{g_{\rho}^2}{2[m_\rho^2 + \Lambda_v (g_{\omega} g_{\rho} \omega_0)^2]} n_0\ , \\
  \label{eq:lsym}
  L_{\rm sym} & = & 3E_0 + \frac{1}{2}\left(\frac{3\pi^2}{2}n_0 \right)^{2/3} \frac{1}{E_{\rm F}} \times \nonumber \\
  & & \left(\frac{g_{\omega}^2}{m_\omega^2 + 3c_3 \omega_0^2}\frac{n_0}{E_{\rm F}}
            - \frac{K_0}{9E_{\rm F}} - \frac{1}{3} \right) \nonumber \\
        & & - 3\left(\frac{g_\rho^2}{m_\rho^2 + \Lambda_v (g_{\omega} g_{\rho} \omega_0)^2}\right)^2 \frac{g_{\omega}^3 \Lambda_v \omega_0 n_0^2}{m_\omega^2 + 3c_3 \omega_0^2}\ , \\
  \label{eq:k0}
  K_0 & = & \frac{3p_{\rm F}^2}{E_{\rm F}} + \frac{3M_{\rm N}^\ast p_{\rm F}}{E_{\rm F}} \frac{dM_{\rm N}^\ast}{dp_{\rm F}} + \frac{9g_{\omega}^2}{m_\omega^2 + 3c_3 \omega_0^2}n_0\ , \\
  \label{eq:j0}
  J_0 & = & 3n_0 \frac{d K(n_{_B})}{d n_{_B}} - 12K_0 \; ,
\end{eqnarray}
where $p_{\rm F}$ denotes the Fermi momentum at saturation density, $E_{\rm F}=\sqrt{p_{\rm F}^2+M_N^{\ast 2}}$, and $K(n_{B})$ represents the density-dependent incompressibility with $K_0 = K(n_0)$. The derivative $dK(n_{B})/dn{_B}$ can be expressed as:
\begin{eqnarray}
\label{eq:dkdn}
  \frac{d K(n_{_B})}{d n_{_B}} & = & \frac{\pi^2}{2} \biggl[\frac{6}{p_{\rm F} E_{\rm F}} - \frac{3p_{\rm F}}{E_{\rm F}^3} - \frac{6M^*_N}{E_{\rm F}^3} \frac{d M^*_N}{d p_{\rm F}} + \frac{3M^*_N}{E_{\rm F} p_{\rm F}^2} \frac{d M^*_N}{d p_{\rm F}} \nonumber \\
  & & + \frac{3}{p_{\rm F} E_{\rm F}} \left(\frac{d M^*_N}{d p_{\rm F}}\right)^2 - \frac{3M_N^{\ast 2}}{E_{\rm F}^3 p_{\rm F}} \left(\frac{d M^*_N}{d p_{\rm F}}\right)^2 \nonumber \\
  & & + \frac{3M^*_N}{p_{\rm F} E_{\rm F}} \frac{d^2 M^*_N}{d p_{\rm F}^2} \biggr] + \frac{9g_\omega^2}{m_\omega^2 + 3c_3 \omega_0^2} \nonumber \\
  & & - \frac{54c_3 \omega_0 g_\omega^3}{(m_\omega^2 + 3c_3 \omega_0^2)^3} n_0\ .
\end{eqnarray}
Although an algebraic expression between nuclear properties and couplings can be obtained when $c_3$
vanishes~\citep{2023PhRvC.108b5809Z}, the presence of a nonzero $c_3$ necessitates the use of a numerical root-finding method to determine the coupling constants from nuclear properties, due to their complex and nonlinear relationships in Eqs.~(\ref{eq:esym})--(\ref{eq:j0}). We implemented the Newton-Raphson method to solve these equations, using the auto-differentiation technique provided by \texttt{JAX} package.

In the following FIM analyses, we must choose fiducial parameters or an injected EoS as the truth to generate the simulated GW signals. We adopted the same parameters and EoS from Ref.~\citep{2025arXiv250811875Z} with RMF parameters: $n_0=0.16\,{\rm fm}^{-3}$, $E_0=-16\,{\rm MeV}$, $K_0=260\,{\rm MeV}$, $J_0=-350\,{\rm MeV}$, $E_{\rm sym}=31\,{\rm MeV}$, $L_{\rm sym}=40\,{\rm MeV}$, $M^*_N/M_N=0.7$.

\subsection{Crustal EoS parameterization}
The RMF model provides a reliable description of the NS core EoS. However, at lower densities, the NS matter enters the crust region, where nuclei coexist with free neutrons and electrons. A fully self-consistent calculation of both core and crust EoS within the RMF framework is computationally expensive and impractical for our purposes, since our analysis requires a large number of EoS. Fortunately, the outer-crust is well established and accurately described by the Baym-Pethick-Sutherland (BPS) model~\citep{1971ApJ...170..299B} or the more recent BSk22 EoS~\citep{2018MNRAS.481.2994P}. Therefore, we only need to parameterize the inner-crust region between the outer crust and core. Accordingly, we adopt a polynomial function for the enthalpy-number density relation~\citep{zhu2025inprep}
\begin{eqnarray}
  \label{eq:ic_param}
  \log h(t) & = & [\log h(1) - \log h(0)]t + \log h(0) \nonumber \\
            &   & + t(1-t)(A_0 + A_1 t + A_2 t^2 ),
\end{eqnarray}
where $t=(\log n - \log n_0)/(\log n_{\rm cc} - \log n_0)$, and $n_{\rm cc}$ denote the number densities at the core-crust interface. Usually, the sound speed is expected to decrease drastically after the inner-outer crust transition due to the onset of the neutron liquid~\citep{Schaffner-Bielich_2020}. For simplicity, we set the sound speed in this decreasing region to zero, and let it start increasing from $n_0$. In principle, by adjusting the width of the vanishing-sound-speed region $n_0-n_{\rm io}$, where $n_{\rm io}$ denotes the density at inner-outer crust interface, and by controlling the sound speed at $n_0$, we can effectively reproduce the effects of the descending sound speed region on EoS and NS. Beyond $n_0$, there is no physical mechanism expected to further reduce the sound speed, so we assume it to increase monotonically. This leads to the following constraints on the coefficients $A_0$, $A_1$, and $A_2$:
\begin{equation}
  \label{eq:iccs_constraints}
  A_1 \in
  \begin{cases}
    \bigl[\tfrac{C}{2} - A_0 - \tfrac{\delta_0}{2},\ \tfrac{C}{2} - A_0 + \tfrac{\delta_0}{2}\bigr],
    & \\
    \hfill \text{if } C>0 \text{ and } A_0 < -\tfrac{\sqrt{3}+1}{2}\,C & \\
    \bigl[\max(\tfrac{C}{2},\ \tfrac{C}{2} - A_0 - \tfrac{\delta_0}{2}),\ \tfrac{C}{2} - A_0 + \tfrac{\delta_0}{2}\bigr],
    & \\
    \hfill \text{if } C<0 \text{ and } A_0 < \tfrac{\sqrt{3}-1}{2}\,C &
  \end{cases}
\end{equation}
where $\delta_0 = \sqrt{4A_0^2 + 4CA_0 - 2C^2}$ and $C=A_1+A_2$.

The additional conditions come from fixing the sound speed at the two boundaries, and can be written as
\begin{eqnarray}
  \log h(1) - \log h(0) + A_0 & = & \nonumber \\
  &&  \hspace{-2.0cm} (\log n_{\rm cc} - \log n_0)c_s^2|_{\rm min},
  \label{eq:cs2_l} \\
  \log h(1) - \log h(0) - A_0 - A_1 - A_2 & = & \nonumber \\
  && \hspace{-2.0cm} (\log n_{\rm cc} - \log n_0)c_s^2|_{\rm cc} \; ,
  \label{eq:cs2_r}
\end{eqnarray}
where $\log h(1) = \log h_{\rm cc}$ and $\log h(0) = \log h_{\rm io}$ due to the continuity of enthalpy. The quanitty
$c_s^2|_{\rm min}$ is a free parameter representing the sound speed at $n_{0}$, which we set to zero for simplicity. The final condition come from the continuity of energy density at $n_{\rm cc}$:
\begin{eqnarray}
\label{eq:e_r}
  e_{\rm cc} & = & n_0 \log(\frac{n_{\rm cc}}{n_0}) \int_{0}^{1} \exp[\log(\frac{n_{\rm cc}}{n_0})t + \log h(t)] dt \nonumber \\
  & & + h_{\rm io}(n_0-n_{\rm io}) + e_{\rm io}.
\end{eqnarray}
where the second term provides the contribution from the region with approximately zero sound speed.
We use the subscript ${\rm io}$ to denote quantities at inner-outer crust interface and ${\rm cc}$ to denote those at the core-crust interface.

Summarily, we use Eqs.~(\ref{eq:cs2_l})--(\ref{eq:e_r}) to determine the parameters $A_0$, $A_1$, and $A_2$ once $n_0$ is specified. However, not all core EoS can be matched with the BPS outer-crust EoS due to the constraint condition Eq.~(\ref{eq:iccs_constraints}). In practice, we discard those EoS that do not satisfy this condition when sampling the nuclear parameters.

\subsection{TOV and tidal equations}
To map the EoS of a neutron star to its mass, radius, and tidal deformability relations, we solve the TOV and tidal equations for a given EoS~\citep{2010PhRvD..81l3016H, 2020PhRvD.102h4058Z}. However, the original form of these equations expressed in terms of the radial coordinate $r$, has an initially unknown boundary and integration domain. This may result in problematic exterior boundary conditions for some EoSs. To overcome this issue, we rewrite the equations in terms of the logarithmic enthalpy, $\log h = \log[(e + p)/\rho]$~\citep{1992ApJ...398..569L, 1998PhRvD..58b4008L, 2008ApJ...677.1216H, 2010PhDT........18P, steil2017structure}
\begin{eqnarray}
  \label{eq:tt_r2}
  \frac{dr^2}{d\log h} & = & \mathcal{K}_{r^2} = -\frac{2r^2(1-2z)}{4\pi r^2 p + z}, \\
  \label{eq:tt_z}
  \frac{dz}{d\log h} & = & \mathcal{K}_{z} = \left(2\pi e - \frac{z}{2r^2}\right)\frac{dr^2}{d\log h}, \\
  \label{eq:tt_h}
  \frac{d^2 H}{d\log h^2} & = & A_H \frac{dH}{d\log h} - B_H H.
\end{eqnarray}
The $\log h$ variables vanishes at vacuum, thereby providing a well-defined stellar surface and exterior boundary conditions $R=r(0)$ and $M=R z(0)$, which represent the radius and mass of the NS, respectively. The energy density, pressure and rest mass density are denoted by $e$, $p$, and $\rho$. The function $H$ denotes the perturbed metric function induced by the external tidal field. It determines the tidal deformability after matching the vacuum solution at the stellar surface~\citep{2020PhRvD.102h4058Z}. The coefficients $A_H$ and $B_H$ are given by
\begin{eqnarray}
\label{eq:tidal_coef}
    A_H & = & \frac{4\pi r^2(1-2z)(e+3p)}{(4\pi r^2 p + z)^2}, \\
    B_H & = & \frac{A_H}{e+3p}\left(4e+8p+(p+e)(1+\kappa) - \frac{3}{2\pi r^2} \right)- 4, \nonumber \\
\end{eqnarray}
where $\kappa=1/c_s^2$. The Eqs.~(\ref{eq:tt_r2})--(\ref{eq:tt_h}) are integrated from $\log h_c$ (central log-enthalpy)
to $0$ to yield radius, mass and tidal deformability.  Because we have assumed the thermodynamic relations
$h=(e+p)/\rho = de/d\rho$ when rewriting these equations, we must ensure that this relation is satisfied in the input EoS table to maintain thermodynamic consistency.

One of the advantage of the enthalpy-based equations~(\ref{eq:tt_r2})--(\ref{eq:tt_h}) is that their derivatives or variations can be expressed analytically due to the well-defined domain and borders. The variational forms of TOV and tidal equations provide an accurate and efficient framework for computing the derivatives of neutron star properties with respect to the central enthalpy or EoS.
These derivatives are particularly important for gradient-based optimization and Fisher matrix analyses. The variational equations are written as~\citep{2025PhRvD.111g4026L}
\begin{eqnarray}
  \label{eq:dtt_r2}
  \frac{d \Delta_{r^2}}{d\log h} & = & \mathcal{K}_{r^2, r^2} \Delta_{r^2} + \mathcal{K}_{r^2, z} \Delta_{z} + \mathcal{K}_{r^2, p} \Delta_{p} \\
  \label{eq:dtt_z}
  \frac{d \Delta_{z}}{d\log h} & = & \mathcal{K}_{z, r^2} \Delta_{r^2} + \mathcal{K}_{z, z} \Delta_{z} + \mathcal{K}_{z, p} \Delta_{p} + \mathcal{K}_{z, e} \Delta_{e}, \\
  \label{eq:dtt_h}
  \frac{d \Delta_{H}}{d\log h} & = & \Delta_{H'}, \\
  \label{eq:dtt_hp}
  \frac{d \Delta_{H'}}{d\log h} & = & A_H \Delta_{H'} - B_H \Delta_{H} + \sum_{i} (A_{H, i} H' - B_{H, i} H) \Delta_{i}, \nonumber \\
\end{eqnarray}
where $\mathcal{K}_{r^2,i}$, $\mathcal{K}_{z,i}$, $A_{H, i}$, and $B_{H, i}$ denote the corresponding derivatives with respect to $i=r^2,z,e,p,\kappa$. The explicit expressions of these terms are listed in Appendix~\ref{sec:appendix1}. Equations~(\ref{eq:dtt_r2})--(\ref{eq:dtt_hp}) describe how the quantities $r^2$, $z$ and $H$ vary in response to the variations of the EoS, which are represented by $\Delta e$, $\Delta p$, and $\Delta \kappa$. These equations can be solved simultaneously with the original TOV and tidal equations once the variational EoS is specified.

If we consider a variation of $\log h$
\begin{eqnarray}
\label{eq:delta_loghc}
  \log h \to \log h + \delta \log h = \log h \left(1 + \frac{\Delta \log h_c}{\log h_c} \right),
\end{eqnarray}
while keeping the EoS fixed, this variation effectively rescales the log-enthalpy by a factor of $(1 + \Delta \log h_c / \log h_c)$ at every grid, which corresponds to a small change in the central log-enthalpy. Applying this variation to the TOV and tidal equations yields the corresponding differentiation with respect to the central log-enthalpy
\begin{eqnarray}
\label{eq:delta_v}
  \Delta \left(\frac{d v}{d \log h}\right) = \frac{d \Delta_v}{d \log h} - \frac{d v}{d \log h},
\end{eqnarray}
where $v$ represents the variables $r^2$, $z$, and $H$. Compare to Eqs.~(\ref{eq:dtt_r2})--(\ref{eq:dtt_hp}), the differentiation equations with respect to the central log-enthalpy include an additional term on the RHS arising from the variation of $\log h$ itself.
\begin{eqnarray}
\label{eq:delta_v2}
  \frac{d \Delta_v}{d \log h} & = & \sum_{i} \mathcal{K}_{v, i} + \frac{d v}{d \log h}.
\end{eqnarray}
In this context, the variations of the EoS quantities become explicitly known
\begin{eqnarray}
  \label{eq:depcs_hc}
  \Delta e = \kappa (e+p) \log h, \\
  \Delta p = (e+p) \log h,  \\
  \Delta \kappa = \frac{d \kappa}{d\log h} \log h.
\end{eqnarray}

By combining the solution of Eqs.~(\ref{eq:dtt_r2})--(\ref{eq:dtt_hp}) and (\ref{eq:delta_v2}), we can compute the derivative of all NS observables (mass, radius, and tidal deformability) with respect to the variations in the EoS. This provides a direct and analytical method for evaluating the derivatives required in the Fisher matrix calculation.

\subsection{Mass distribution and BNS merger population}

In order to estimate the uncertainties of EoS constrained by 3G detectors, we have to generate simulated GW signals from
the BNS mergers. The signals that can be detected depend on the sensitivity of the detectors, their redshifts, BNS
merger event rates, observing time and the masses of NSs. In this work, we consider a network of two 3G detectors, ET and CE1~\citep{2010CQGra..27h4007P, 2012CQGra..29l4013S, 2020JCAP...03..050M, 2019BAAS...51g..35R}, whose sensitivity curves are adopted from
the \texttt{GWFast} package~\citep{2022ApJ...941..208I, 2022ApJS..263....2I}. 

The differential BNS merger rate as function of redshift $R(z)$ can be written in terms of the volumetric BNS merger rate
density $\mathcal{R}(z)$ in the source frame as
\begin{eqnarray}
  \label{eq:dRdz}
  \frac{dR(z)}{dz} & = & \frac{dV_c}{dz} \frac{1}{1+z} \mathcal{R}(z),
\end{eqnarray}
where $dV_c/dz$ is the differential comoving volume, and the factor $1/(1+z)$ accounts for the time dilation between the
source frame and detector frame. In general, $\mathcal{R}(z)$ is a function of redshift, and can depend on the star
formation rate and the delay time between binary formation and its coalescence~\citep{2023JCAP...07..068B,
  2024A&A...681A..56S, 2025MNRAS.541..798D, 2025arXiv250720258S}. For simplicity, we conservatively assume a constant merger rate density $\mathcal{R}(z)=p$ in this work, and vary the value of $p$ to represent different population scenarios. The total number of events are obtained by integrating Eq.~(\ref{eq:dRdz}) over redshift up to $z=5$ and times the observing time, we set the threshold for detection to be ${\rm SNR} > 8$.

We adopt the Planck 2018 model for the relation between luminosity distance and redshift. Redshifts are drawn up to $z=5$ from the normalized distribution proportional to $(dV_c/dz)/(1+z)$. The waveform is evaluated using the detector-frame chirp mass $\mathcal{M}_{\rm det}=(1+z)\mathcal{M}_{\rm src}$. This degeneracy enters the Fisher matrix through the derivatives of waveform with respect to the chirp mass and luminosity distance, and ultimately increase the uncertainties in the EoS parameters, particularly for the low SNR events.

The BNS merger rate $p$ remains highly uncertain based on current observations~\citep{2023PhRvX..13a1048A, 2026ApJ..1005L..51A, 2026arXiv260527226T}, with estimates ranging from $\sim 5.1$ to $155\ {\rm Gpc}^{-3},{\rm yr}^{-1}$ according to GWTC-5.0 data. We therefore consider two cases for our simulated BNS samples:
(1) a conservative scenario with a lower population of $p=100\ {\rm Gpc}^{-3},{\rm yr}^{-1}$ accumulated over one year of observation, including $\sim$ 25,000 detectable events (${\rm SNR} > 8$); and
(2) an optimistic scenario with 10 times more observed events of the conservative scenario ($\sim$ 250,000 events with ${\rm SNR} > 8$). This scenario corresponds to larger merger event rate and longer observing time, representing the best achievable precision for 3G detectors.
Both scenarios comfortably underestimate the full potential of 3G networks, since the star formation and thus merger rate increases
strongly with redshift.

\begin{figure*}[t!]
  {\centering
  \includegraphics[width=0.495\textwidth]{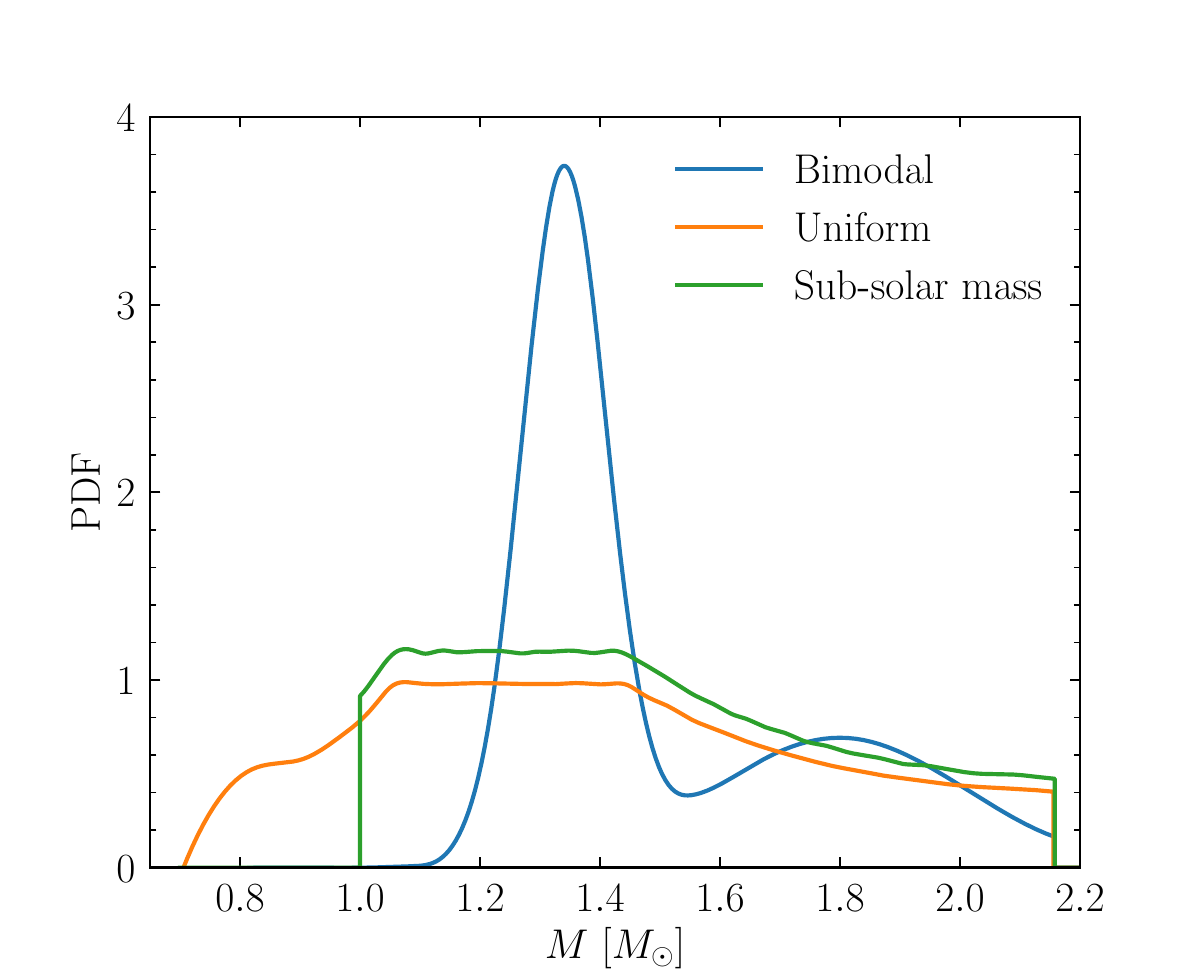}
  \includegraphics[width=0.495\textwidth]{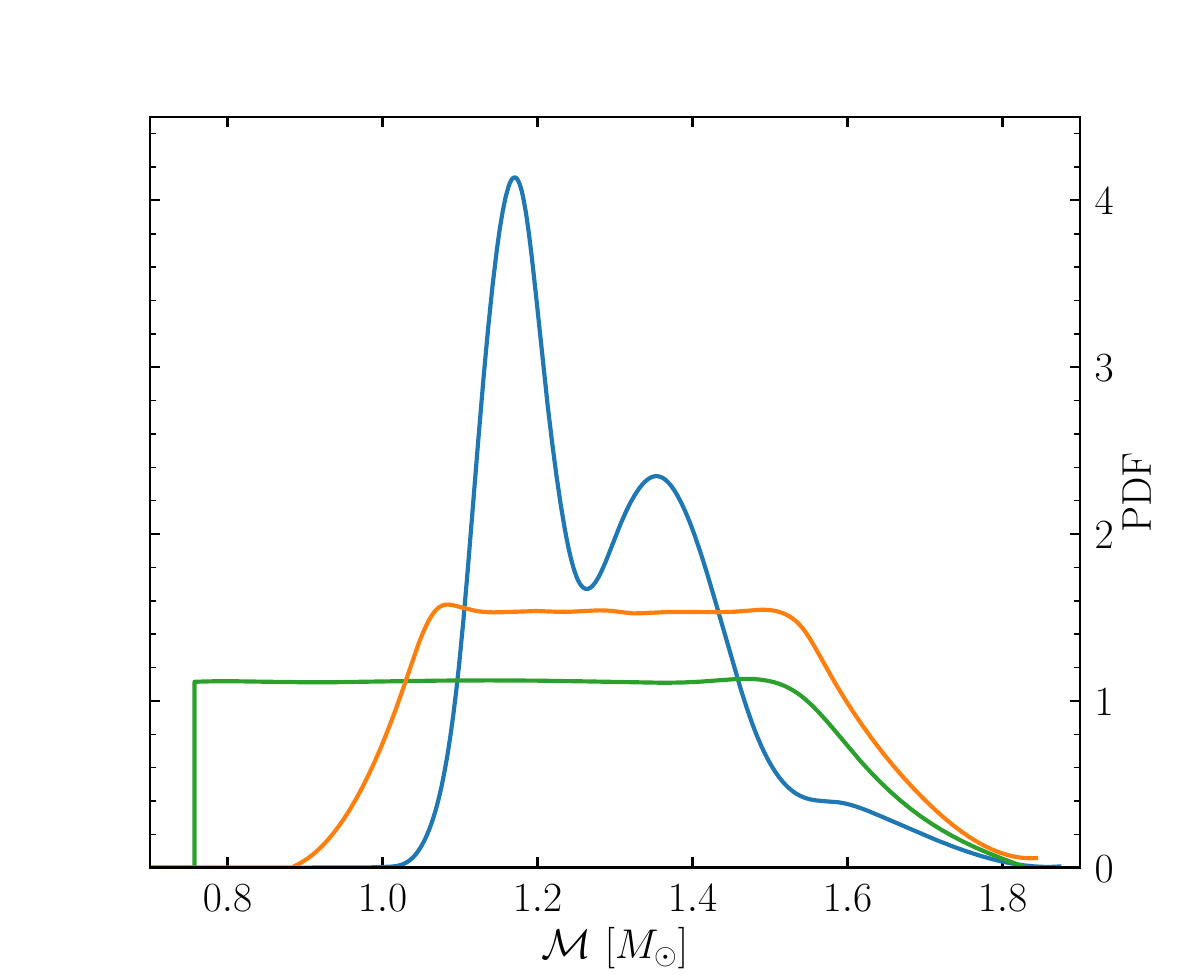}}
  \caption{The NS mass (left) and chirp mass (right) distributions for the BNS systems. The blue line denotes the realistic bimodal Gaussian distribution given by Eq.~(\ref{eq:mass_dist}), while the orange and green lines represent the uniform distribution excluding and including sub-solar mass NSs, respectively. Due to the cutoff of maximum NS mass, the chirp mass distribution for the uniform distribution is not constant anymore over the range $[1.5, 1.8]\,\Msun$.}
  \label{fig:m_dist}
\end{figure*}

The masses of NS in BNS systems play a important role in determining both the detectability of GW signals and the constraints on the EoS. Massive NSs generally produce stronger GW signals and are therefore easier to detect. However, they are more difficult to be tidally deformed, leading to weaker matter effects in the waveform and, consequently, transmit less information about the EoS. Observational data suggest that the NS mass distribution is bimodal~\citep{2018MNRAS.478.1377A}, though it remains uncertain whether the same distribution applies to NSs in BNS systems. Moreover, the component masses within BNS systems may not be statistically independent.
On the other hand, a light NS has been reported within the supernova remnant HESS J1731-347~\citep{2022NatAs...6.1444D}, suggesting the possible existence of sub-solar mass NSs. The formation mechanisms of them have been discussed in Ref.~\citep{2024ApJ...971L..34M, 2025ApJ...991L..22C}, and their potential GW signals have been searched for in the recent LIGO-Virgo-KAGRA (LVK) Observing Runs~\citep{2023MNRAS.524.5984L, 2026arXiv260505444T}. Nevertheless, sub-solar mass NSs are typically not included in standard mass distributions due to their rarity.

To investigate the impact of different mass distributions on the EoS constraints encompassing these observational uncertainties, we consider two mass-distribution models. The first is a realistic bimodal distribution given by
\begin{eqnarray}
\label{eq:mass_dist}
P(m) & = & \sum_{i=1,2} r_i \mathcal{N}(m; \mu_i, \sigma_i) \Theta(m_{\rm max} - m) / \Phi_i,
\end{eqnarray}
where $\mathcal{N}(m; \mu_i, \sigma_i)$ denotes a Gaussian distribution with mean $\mu_i$ and standard deviation
$\sigma_i$, and $\Theta$ is the step function introducing a cutoff at $m_{\rm max}=2.17,\Msun$, corresponding to the NS
maximum mass for our injected EoS. The weights $r_1=0.65$, $r_2=1-r_1$, and normalization factors $\Phi_i$ ensure that $P(m)$ is normalized to unity. We adopt the best-fit parameters $\mu_1=1.34$, $\sigma_1=0.07$, and $\mu_2=1.80$, $\sigma_2=0.21$ from Ref.~\citep{2018MNRAS.478.1377A}. We assume that the component masses in BNS systems are independent, \ie the masses of both NSs in binary systems are drawn from the same distribution Eq.~(\ref{eq:mass_dist}).

The second model assumes a uniform distribution in chirp mass $\mathcal{M} \in [0.75, 1.8],\Msun$ and symmetric mass ratio $\eta \in [0.24, 0.25]$, which includes sub-solar mass NSs. To directly compare this with the realistic bimodal distribution Eq.~(\ref{eq:mass_dist}) and evaluate the impact of potential sub-solar mass NSs, we perform analyses both including and excluding these sub-solar mass samples.

In Fig.~\ref{fig:m_dist}, we present the mass and chirp mass distributions of BNS systems for the three models. Note that the  probability density function (PDF) of $\mathcal{M}$ for uniform distribution (orange and green lines) is not constant over the range $[1.5, 1.8]\,\Msun$, due to the sharp cut-off of the NS maximum mass. For the uniform distribution model excluding sub-solar mass NSs (green line), BNS samples containing one or more sub-solar mass components ($<1.0\,\Msun$) are discarded from the analysis.

The simulated GW signals are generated using the \texttt{IMRPhenomD\_NRTidalv2} waveform
model~\citep{2016PhRvD..93d4007K, 2019PhRvD.100d4003D}. The tidal deformability of each NS is computed from its mass and
the injected EoS.
The spins of NSs are neglected in this study, as they have negligible impact on the EoS constraints under the low-spin scenario.

\subsection{Fisher information matrix}

To compute the posterior distributions of NS and EoS parameters from the simulated GW signals with respect to the detector network (ET+CE1), we employ the Fisher information matrix (FIM) method~\citep{1994PhRvD..49.2658C,1995PhRvD..52..848P,2008PhRvD..77d2001V}. The FIM is computed as
\begin{eqnarray}
  \label{eq:fisher}
  F_{ij} & = & \sum_{\rm d} F_{ij}^{\rm d} = \sum_{\rm d} \left< \frac{\partial h^{\rm d}(f, \vec{\theta})}{\partial \theta_i} \Big| \frac{\partial h^{\rm d}(f, \vec{\theta})}{\partial \theta_j} \right>_{\rm d}.
\end{eqnarray}
where $h^{\rm d}(f, \vec{\theta})$ denotes the GW strain in the frequency domain for detector $d$, and $\vec{\theta}$ represents the set of parameters describing the GW signal. In this work, we are primarily interested in the EoS parameters. Consistent with the implementation described above, all remaining waveform parameters are held fixed at their injected values when the EoS-only Fisher matrix is extracted. The inner product $\left<a|b\right>$ is defined as
\begin{eqnarray}
  \label{eq:inner_product}
  \left<a|b\right>_{\rm d} & = & 2\int_{-\infty}^{\infty} \frac{a^\ast(f)b(f) + a(f)b^\ast(f)}{S_n^{\rm d}(|f|)} df,
\end{eqnarray}
where $S_n^{\rm d}(f)$ denotes the one-sided noise power spectral density of detector $d$.

We compute the FIM for each simulated BNS event using the \texttt{GWFast} package~\citep{2022ApJ...941..208I, 2022ApJS..263....2I}. In our analyses, we retain only those events with SNR larger than 8, for which the linearized-signal approximation should be sufficiently accurate to justify the use of the FIM method~\citep{2008PhRvD..77d2001V}. Furthermore, we assume that all NSs share the same EoS, allowing us to accumulate the FIM of EoS parameters $\cal{Y} \in (K_0, J_0, E_{\rm sym}, L_{\rm sym}, M^*_N)$ over all detected events. Note that we fixed the saturation density $n_0$ and binding energy $E_0$ in our analyses, as they are well constrained by nuclear experiments compared with other parameters, and have negligible impact on the EoS within their current constraints. 

For the EoS parameters $\cal{Y}$, our current knowledge from nuclear experiments and astrophysical observations can further be incorporated as prior information by introducing a prior FIM. We adopt a diagonal prior FIM with standard deviations of $3\,{\rm MeV}$, $30\,{\rm MeV}$, $40\,{\rm MeV}$, $300\,{\rm MeV}$, and $0.1$ for $E_{\rm sym}$, $L_{\rm sym}$, $K_0$, $J_0$, and $M^*_N$, respectively~\citep{2025arXiv250811875Z}. Although its contribution is negligible compared to the total FIM of all events, the prior FIM helps to regularize the matrix inversion and prevent singularities, particularly when the number of events is small. More details can be found in Ref.~\citep{2025arXiv250811875Z}. In Appendix~\ref{sec:appendix5}, we quantify how projected FRIB-era laboratory constraints on the symmetry-energy sector, encoded as correlated prior-FIM blocks, would sharpen these forecasts.

\subsection{Simulated data}
Having established the mass distribution and BNS merger population, we proceed to generate simulated GW signals from BNS
mergers.
   For each event, we randomly draw the component masses from the specified mass distribution model, and assign random sky locations and orientations. The redshift or luminosity distance of each event is determined based on the assumed spatially uniform distribution and the merger rate. The details of all simulated data groups used in this paper are listed in Table~\ref{tab:sim_data}.

\begin{table}[t]
\begin{center}
\begin{tabular}{cccc}
\hline
Name &  Distribution & No. of events\\
     &               & (SNR$>8$) \\
\hline
\texttt{B2} & \multirow{2}{*}{$m1,m2 \in$ Eq.~(\ref{eq:mass_dist})} & $\sim$ 25,000 \\
\texttt{B3} &  & $\sim$ 250,000 \\
\hline 
\texttt{U2} & \multirow{2}{*}{\shortstack{$\mathcal{M} \in [0.75, 1.8],\ \eta \in [0.24, 0.25]$ \\ without s.s.}} & $\sim$ 25,000 \\
\texttt{U3} &  & $\sim$ 250,000 \\
\hline 
\texttt{S2} & \multirow{2}{*}{$\mathcal{M} \in [0.75, 1.8],\ \eta \in [0.24, 0.25]$} & $\sim$ 25,000 \\
\texttt{S3} &  & $\sim$ 250,000 \\
\hline
\end{tabular}
\end{center}
\caption{Details of simulated BNS merger GW data used in this work. We use three different mass distribution models and two different number of events to generate six groups of simulated data. The labels \texttt{B}, \texttt{U}, and \texttt{S} in the group names represent the bimodal Gaussian distribution given by Eq.~(\ref{eq:mass_dist}), uniform distribution in chirp mass and symmetric mass ratio after removal of sub-solar mass NSs, and uniform distribution including sub-solar mass NSs, respectively. The numbers 2 and 3 in the name denote cases of low (100 ${\rm Gpc}^{-3}\ {\rm yr}^{-1}$) rate over 1 year of observation and 10 times larger number of events ($\sim 250,000$) than previous case, respectively.
}
\label{tab:sim_data}
\end{table}

We generated six groups of simulated data using three different mass distribution models and two different events number, as summarized in Table~\ref{tab:sim_data}. The realistic bimodal Gaussian distribution given by Eq.~(\ref{eq:mass_dist}), the uniform distribution in chirp mass and symmetric mass ratio, excluding sub-solar mass NSs, and the same uniform distribution but including sub-solar mass NSs are denoted as \texttt{B}, \texttt{U}, and \texttt{S}, respectively. For each model, we consider two events number scenarios: a merger event rate of 100 ${\rm Gpc}^{-3}\ {\rm yr}^{-1}$ over one year of observation (denoted by 2), corresponding to $\sim 25,000$ events, and a larger number of events ($\sim 250,000$) which is 10 times than previous scenario (denoted by 3).

\begin{figure*}[t!]
  {\centering
  \includegraphics[width=0.33\textwidth]{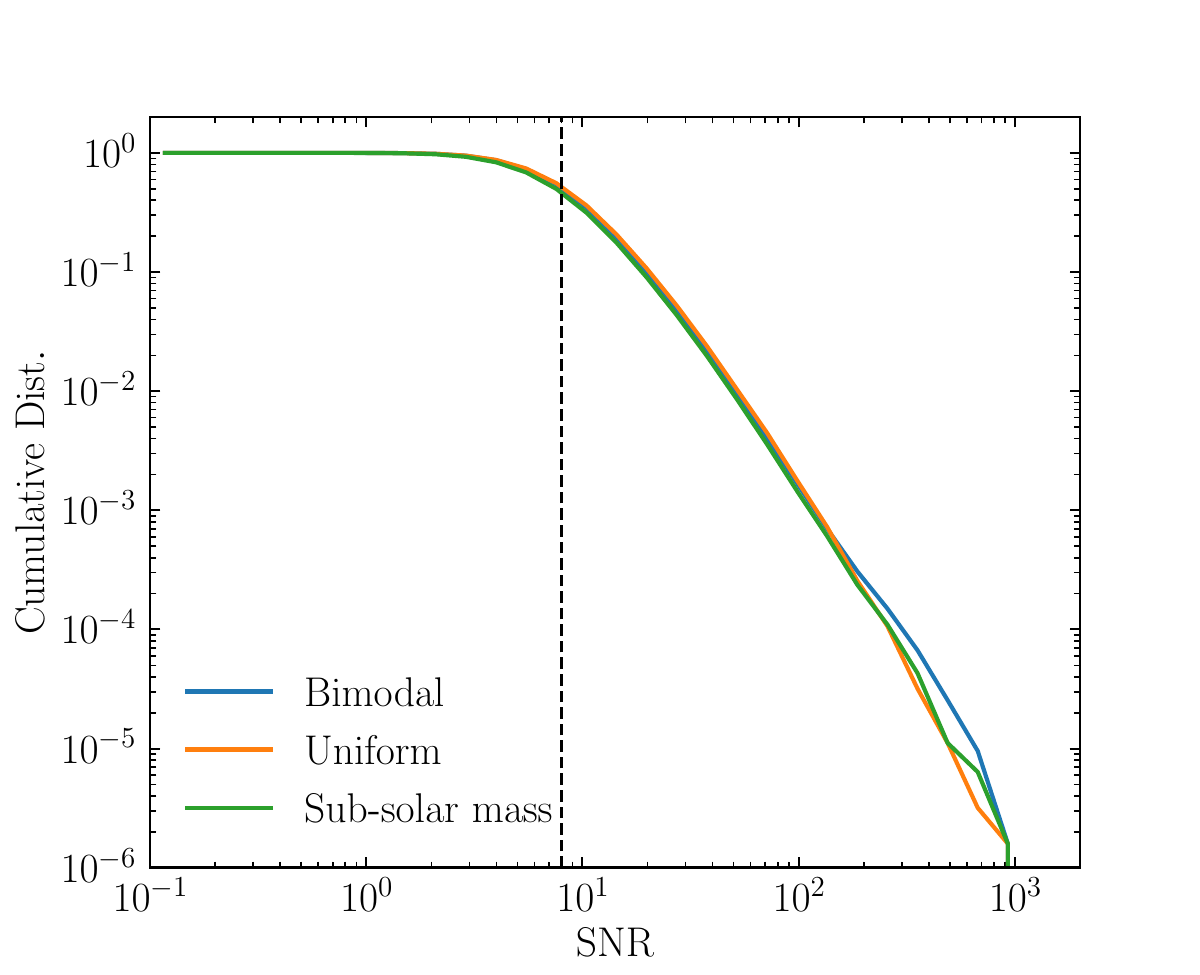}
  \includegraphics[width=0.33\textwidth]{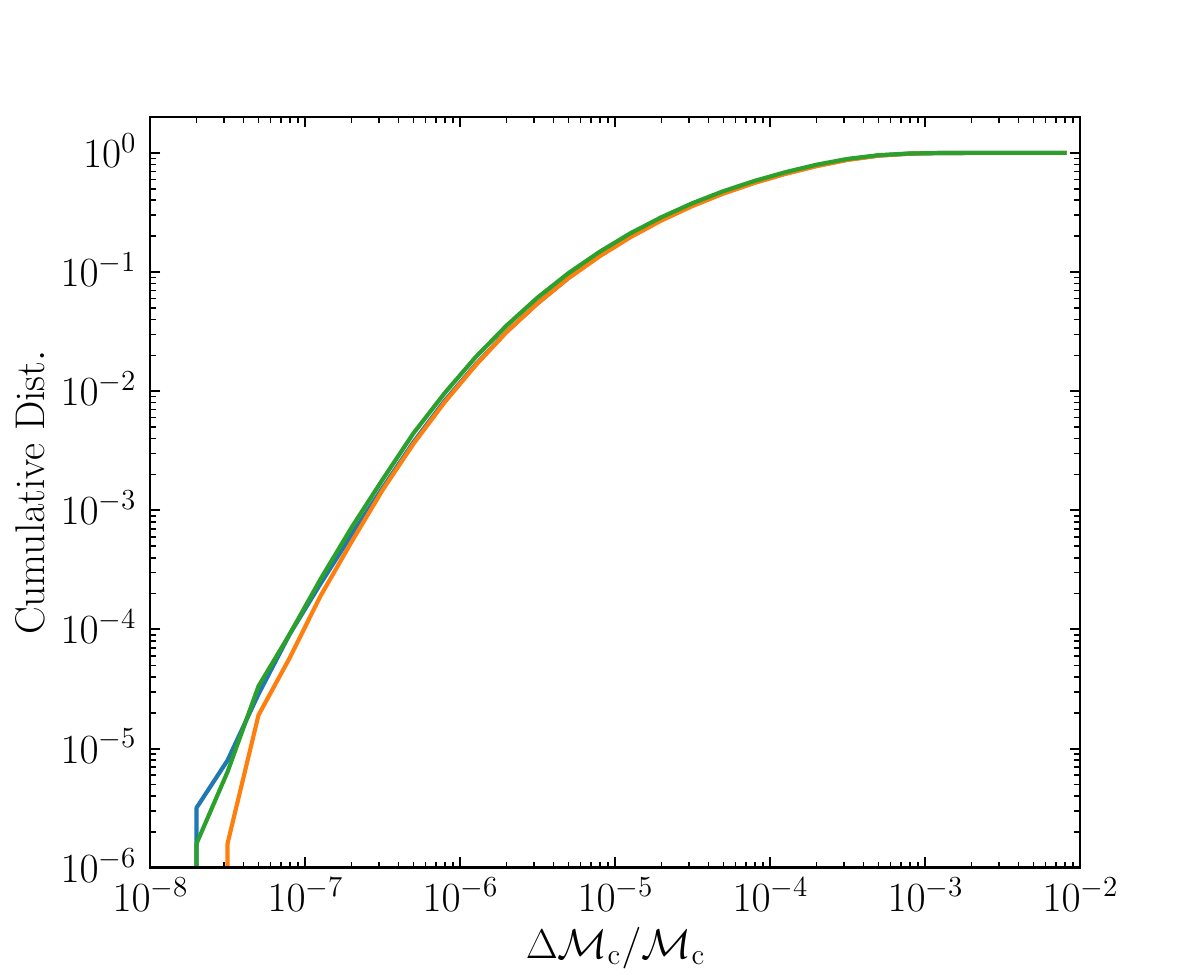}
  \includegraphics[width=0.33\textwidth]{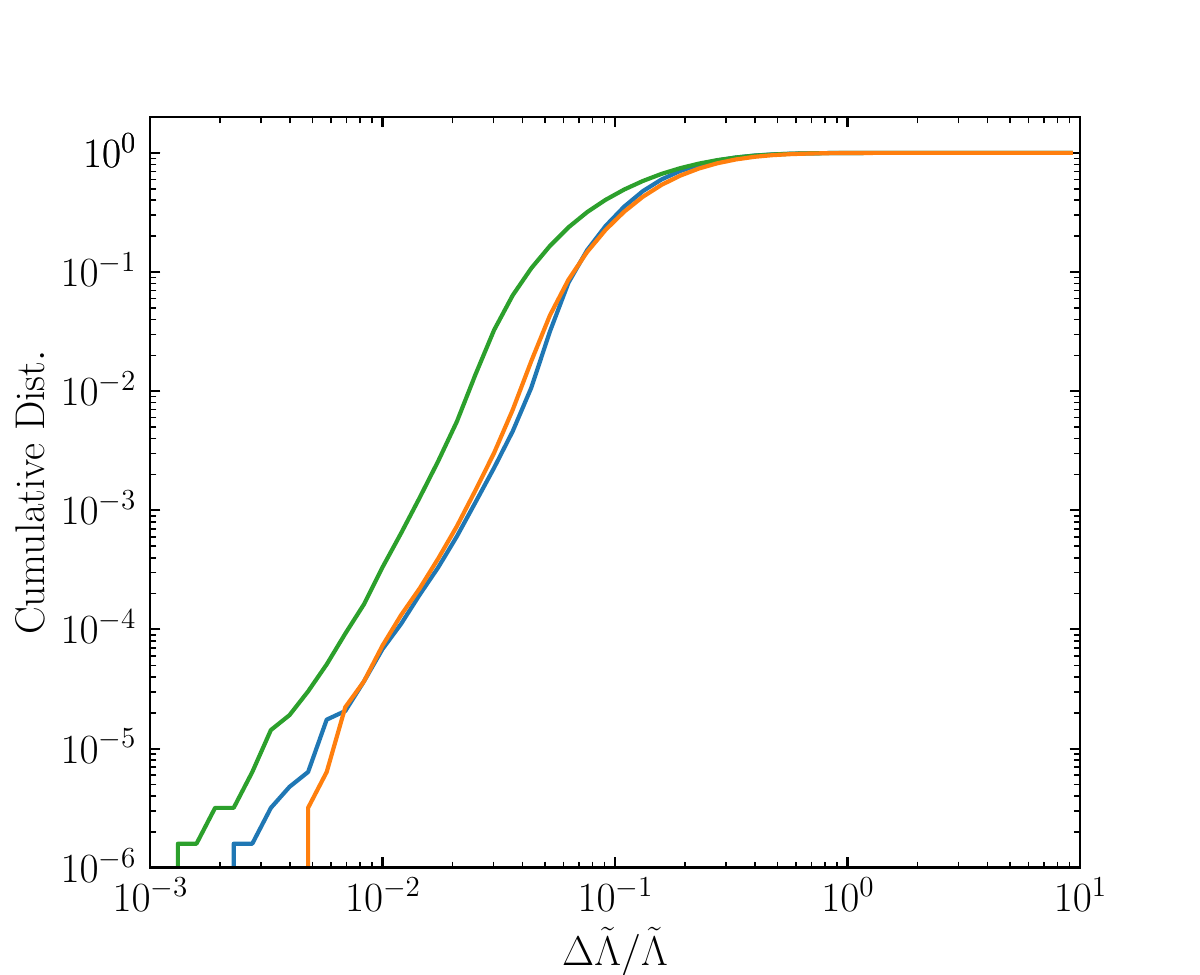}}
  \caption{Cumulative distributions of SNR (left), chirp mass uncertainty (middle), and effective tidal-deformability uncertainty (right) for three different mass distribution models. The bimodal Gaussian distribution, the uniform distribution in chirp mass and symmetric mass ratio excluding sub-solar mass NSs, and the same uniform distribution but including sub-solar mass NSs are represented by the blue, orange, and green lines, respectively. Note that the populations only contribute a factor of the number of events, but does not affect the shapes of the distribution densities. The cumulative distributions are independent of the number of events.}
  \label{fig:snr_dist}
\end{figure*}

In Fig.~\ref{fig:snr_dist}, we present the cumulative distributions of SNR (left), chirp mass uncertainty (middle), and effective tidal-deformability ($\tilde{\Lambda}$) uncertainty (right) for the three different mass distribution models. The populations differ only by an overall factor proportional to the number of events, which does not alter the shape of the distribution densities. Therefore, we only show the results from the high events number case as representative examples. The black vertical dash lines in left panel denotes the SNR threshold of 8 in our analyses.

The SNR and chirp mass uncertainty distributions are similar across all mass distribution models, as they are primarily
determined by the distance. The small differences between them at the high-SNR and low-uncertainty of chirp mass ends
can be attributed to the random sampling fluctuations and do not imply any statistically meaningful differences. In
contrast, the distributions of $\tilde{\Lambda}$ uncertainties exhibit a clear deviation of the sub-solar mass model
from the other two models. Because sub-solar-mass NSs have much larger tidal deformabilities, and therefore induce
stronger matter effects in the waveform, their tidal contributions can be measured more accurately. This leads to larger
fraction of events with precise measurements of tidal deformability $\tilde{\Lambda}$.

\subsection{Hierarchical EOS inference}
In general, the posterior distribution for EOS hyperparameters ${\cal Y}$ follows by combining all the single-event
observations into \cite{2022hgwa.bookE..45V}
\begin{align}
  \label{eq:hierarchical}
  {\cal L}_{\rm net}({\cal Y}) p({\cal Y}) &= p({\cal Y}) \prod_k
  \frac{
    \int d \theta \ell_k(\theta)p(\theta|{\cal P})
  }{\xi({\cal P}) }
\end{align}
where $p({\cal Y})$ is our EOS prior distribution; where $\ell_k$ are single-event parameter likelihoods, here approximated
using the Fisher matrix; and 
where the factor $\xi({\cal P})$, the fraction of detectable sources due to the binary mass distribution model
$p(\theta|{\cal P})$, accounts for detector selection effects~\citep{2022arXiv220400461E, 2023PhRvD.108d3011E}.
 This expression marginalizes over the assumed-unknown
number of merger events of binary neutron stars.

In this work, for simplicity we will assume the mass distribution model $p(\theta|{\cal P})$ is fixed, known, and
completely independent of the assumed EOS parameters ${\cal Y}$. In this fashion, we completely separate knowledge of
the maximum NS mass, conceivably inferred from the NS mass distribution, from direct measurements of NS tidal
deformability versus mass.  In effect, we are conservatively assuming that the compact object mass distribution does
not have a feature precisely and unambiguously associated with the NS maximum mass, and that for simplicity GW measurements do not effectively differentiate
between BH- and NS-containing binaries near the NS maximum mass~\citep{2020PhRvD.102b3025F, 2020ApJ...893L..41C,2024PhRvD.110d3013W}.  In Appendix 
\ref{sec:appendix3} we describe how little knowledge and inference of the NS maximum mass constraints from population
impacts our conclusions.
Because we fix the NS population within each calculation, the factor $\xi(P)$ is constant and does not influence our results.

Under these assumptions, since the population model is fixed, the log-likelihood is approximated as a sum of
single-event quadratic terms. In the numerical implementation, we extract the EoS block of each full waveform Fisher
matrix while holding all non-EoS parameters, including the luminosity distance, fixed at their injected values. We also
omit the contributions associated with our assumed-constant parameter distribution (i.e., $\propto \prod_k
p(\theta_k)/\sqrt{|F_k|}$). This conditional treatment isolates the information carried by the tidal response but is
more optimistic than a joint inference in which masses, distance, redshift, and EoS parameters are simultaneously varied.
As a result, the conditional distribution for ${\cal Y}$ is a normal distribution, whose inverse covariance follows by adding the
EoS-only FIM obtained for each event.

\section{Results}
\label{sec:results}


\subsection{The fiducial model}

We use the model $\texttt{B3}$ as our fiducial case to illustrate the impact of different number of events and mass distribution models on the final EoS constraints from 3G detectors. This model adopts the realistic bimodal Gaussian mass distribution given by Eq.~(\ref{eq:mass_dist}) and a high BNS merger events. This scenario represents the optimistic case for 3G detectors to constrain the EoS and reflects the limiting precision that GW observations alone can achieve. This fiducial model is used and discussed in details in Ref.~\citep{2025arXiv250811875Z}.

\begin{figure*}[t!]
  {\centering
  \includegraphics[width=0.33\textwidth]{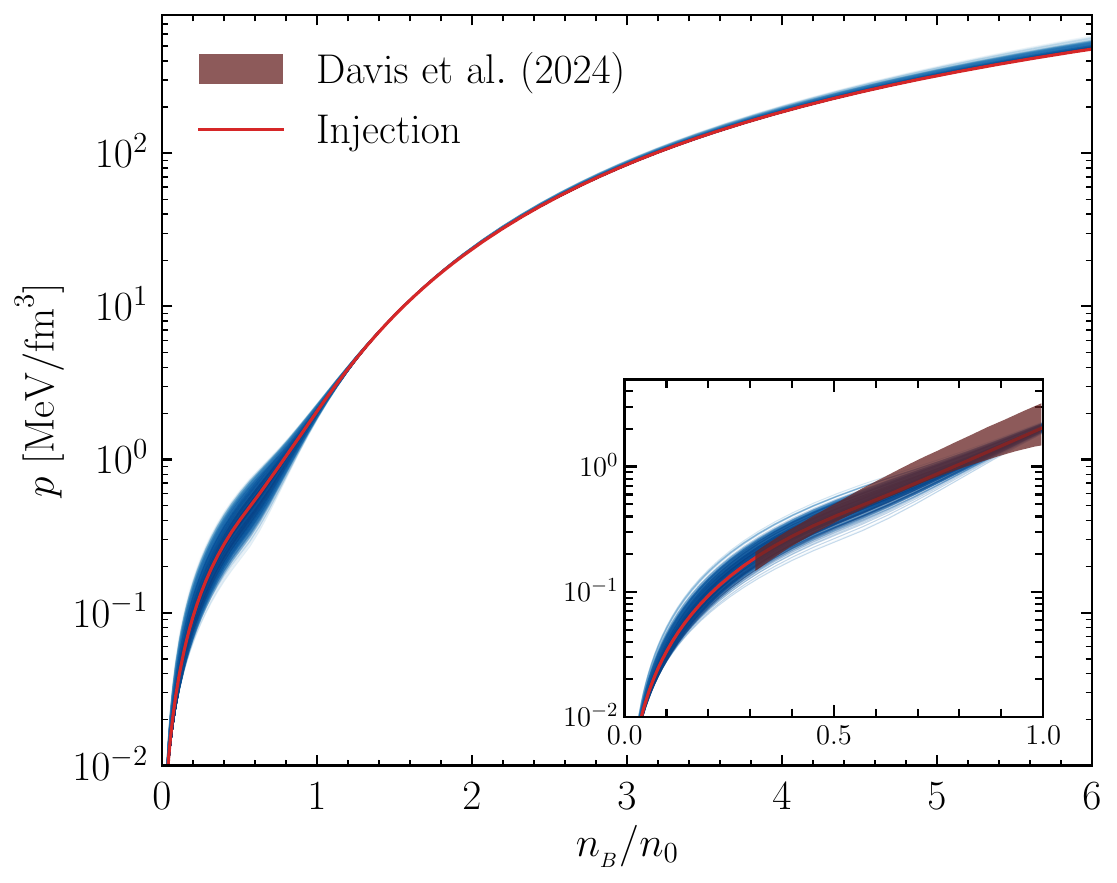}
  \includegraphics[width=0.33\textwidth]{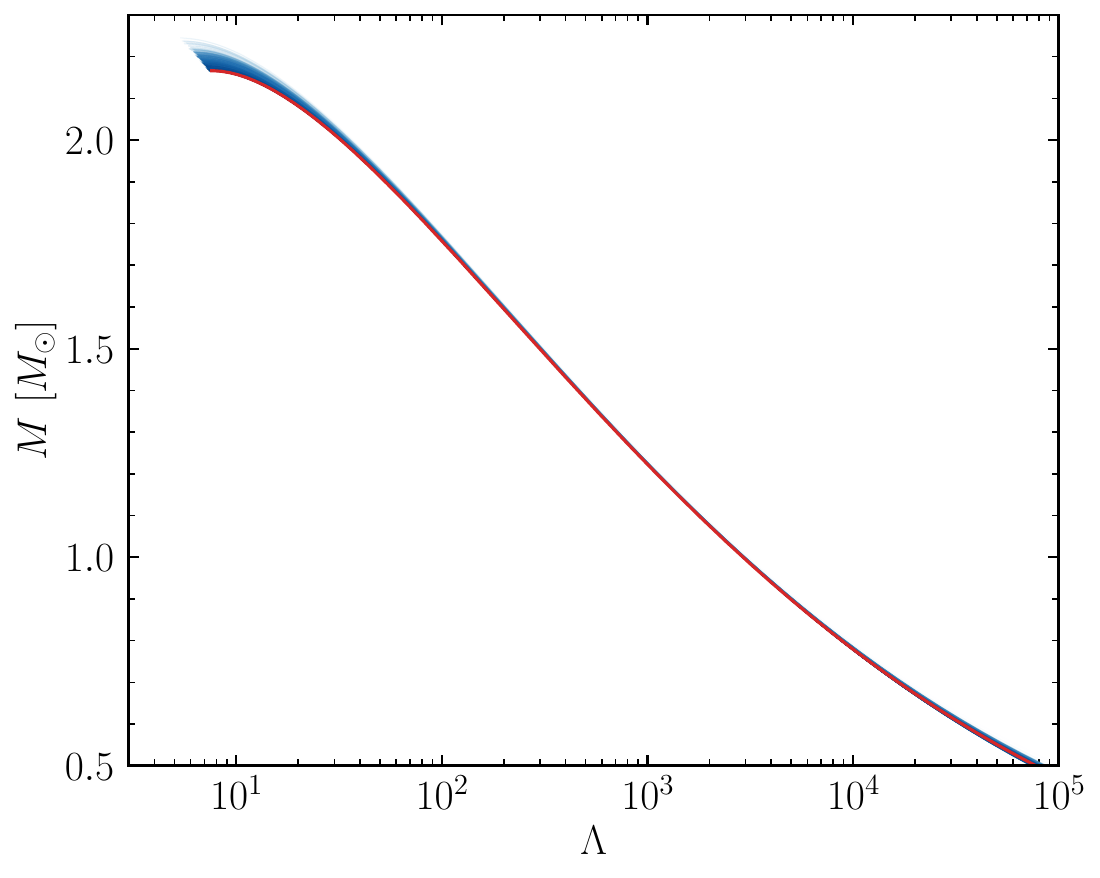}
  \raisebox{-0.07cm}{\includegraphics[width=0.33\textwidth]{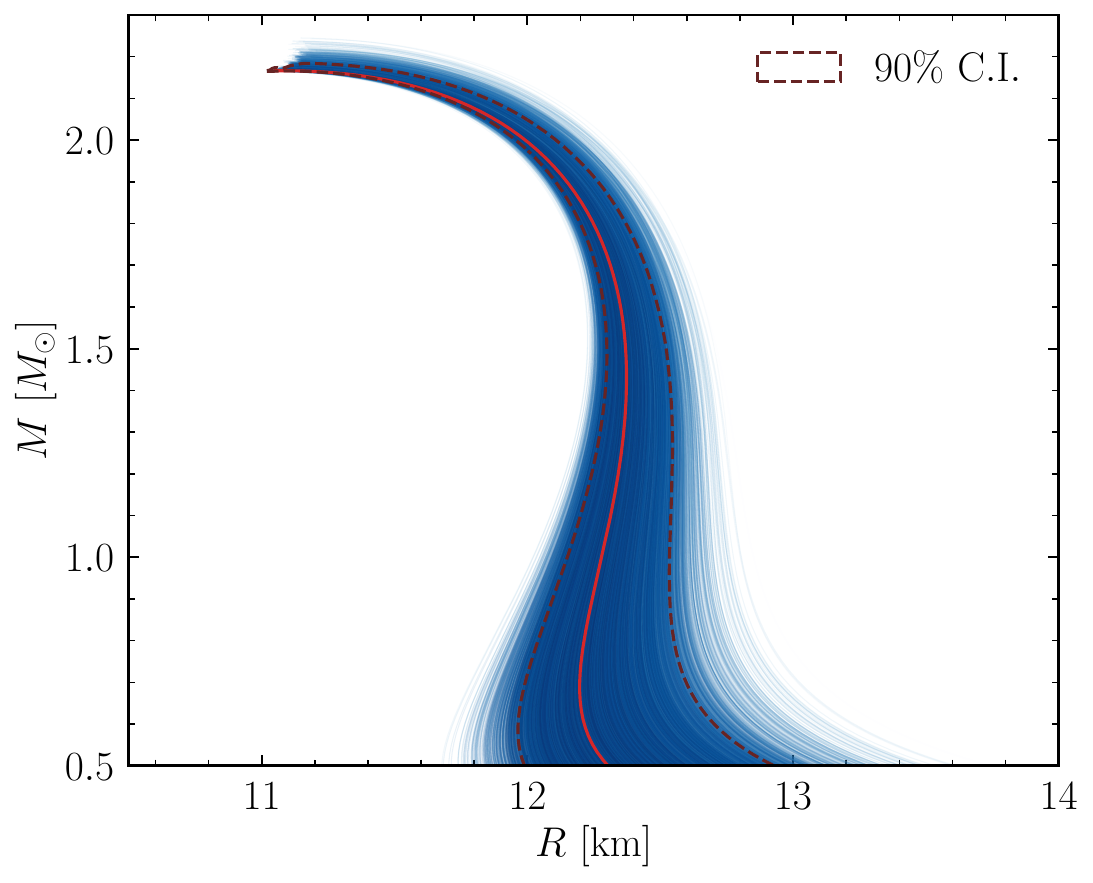}}}
  \caption{The posterior samples of the fiducial \texttt{B3} model (blue lines) and the injection (red) are displayed. The EoS, tidal deformability and M-R relation are shown in the left, middle and right panels, respectively. In the inset of the EoS panel, we also show the current sub-saturation EoS constraints from Ref.~\citep{2024A&A...687A..44D} for comparison, which are denoted by the dark red shaded region. In the right M-R panel, the $90\%$ credible interval of the posterior distribution is also displayed and represented by dash line.}
  \label{fig:b3}
\end{figure*}

Figure \ref{fig:b3} shows the inferred zero-temperature beta-equilibrium EoS (left panel), tidal deformability versus
mass (center panel), and mass-radius distributions.  This figure demonstrates that  3G detectors can constrain two of
the  three observable properties of the EoS model to a very high precision. Although mild uncertainties remain in the crust and sub-saturation
density region of the EoS and in NS radii, due to the rarity of low-mass NSs in the population, they can be
significantly reduced by incorporating the knowledge from the calculation of Chiral Effective Field Theory and nuclear
experiments~\citep{2016PhRvC..93e4314D, 2024A&A...687A..44D, 2024PhRvD.109j3015S, 2025arXiv250516929K}.
The left panel of  Fig.~\ref{fig:b3} has an inset, providing 
a direct comparison between our EoS posterior and  current sub-saturation EoS constraints from Ref.~\citep{2024A&A...687A..44D}, where a unified crust-core EoS model is constructed. Notably, the posterior uncertainties at sub-saturation densities from the 3G detectors are even broader than the present constraints from nuclear physics and observations. This implies that the low-density EoS can further be constrained by combining the information from nuclear physics, see Appendix~\ref{sec:appendix4} for a detailed discussion.
On the other hand, tidal effects weaken for massive NSs, leading to relatively larger uncertainties in the mass-radius relation near the maximum mass. Independent constraints on the maximum mass from NS mass distribution measurements~\citep{2025arXiv251008832M, 2024PhRvD.109d3052F} and multi-messenger observations~\citep{2017ApJ...850L..19M, 2017ApJ...850L..34B} can further improve the EoS constraints.

\subsubsection{Which events are most informative and why?}

The net EoS posterior's inverse covariance  builds up additively from individual source contributions.  In this section, we outline
how individual sources contribute to the overall posterior, highlighting what sources are most informative.

\begin{figure*}[t!]
  {\centering
  \includegraphics[width=1.0\textwidth]{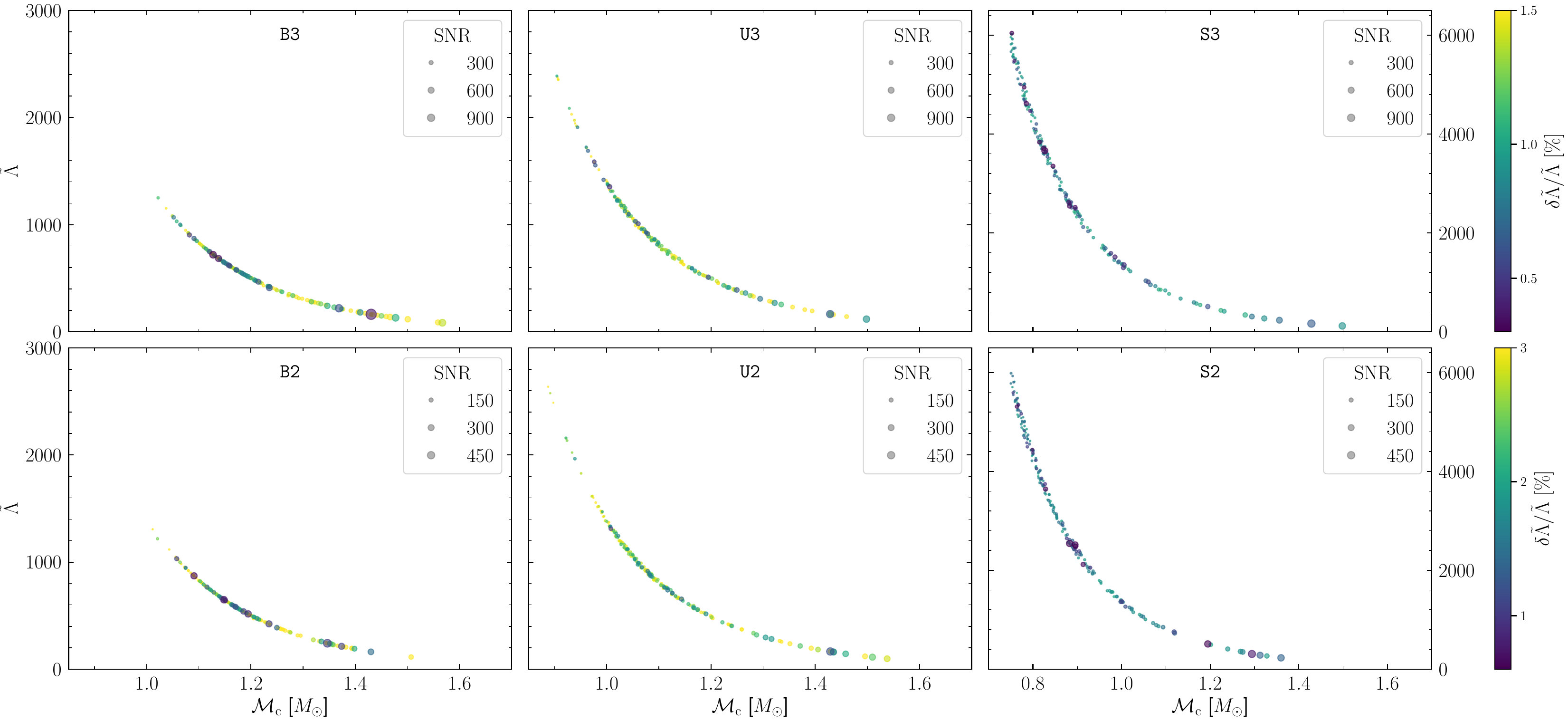}}
  \caption{The best 200 events are presented for each model in the chirp mass-effective tidal deformability plane. The size of each dot represents the SNR of the event, while the color indicates the relative uncertainty of the effective tidal deformability $\delta \tilde{\Lambda} / \tilde{\Lambda}$.}
  \label{fig:mchi_lamtilde}
\end{figure*}

The total assumed number of events directly scales our overall posteriors (i.e., the net inverse covariance matrix is proportional to
$N$, so the precision of individual observables scales as $1/\sqrt{N}$), so though we have adopted a specific nominal number of merger events, our results can be easily scaled to other assumed events number.
Specifically, since the distributions of SNR, chirp mass and tidal deformability remain unchanged, the net Fisher matrix
with a $10\times $ larger number of events is simply a factor of $\sim 10$ larger than that of low events number case
(contributions from prior are minor).   On the other hand, observing more events increases the likelihood of having more high-SNR detections. This is illustrated in Fig.~\ref{fig:mchi_lamtilde}, where a smaller SNR scale is used for the low number of events case. The SNRs in the upper panel are generally larger than those in the lower panel.

Each individual source's contribution also reflects a balance between signal strength (more massive sources are easier to
find) and TD magnitude (low mass NS have larger $\Lambda$).  In Fig.~\ref{fig:mchi_lamtilde}, we present the effective tidal deformability and chirp mass for the 200 events with the best TD measurements for different models. The size of each dot represents the SNR of the event, while the color indicates the relative uncertainty of the effective tidal deformability $\delta \tilde{\Lambda} / \tilde{\Lambda}$. In general, lower mass NSs have larger value of TD and stronger matter effects in the waveform, leading to more accurate TD measurements. Accordingly, we observe the trend that smaller chirp masses correspond to lower TD uncertainties. On the other hand, low NS masses also imply weaker GW signals and lower SNRs. As a result, the best TD measurement events arise from a balance between mass and tidal deformability, we therefore see that most of the low uncertainty dots are concentrated at the intermediate chirp mass region, \ie $1.0$-$1.2\,\Msun$ for both bimodal and uniform models.

Figure ~\ref{fig:dlambda} quantifies the trade-off between signal strength and magnitude of TD, showing the smallest
relative uncertainties for NSs with masses between $1.1$-$1.5\,\Msun$ in Fig.~\ref{fig:dlambda}. This mass range represents the most easily observed mass region after balancing the contribution of mass and TD. At the same time, this range also coincides with the first peak of the bimodal mass distribution, which further decreases the TD uncertainties for the bimodal models. The uncertainties of TD for massive NSs above $2.0\,\Msun$ increase rapidly due to the weakening of tidal effects and the scarcity of such massive NSs in the population. Meanwhile, when comparing the TD uncertainty at low mass end, where the prior knowledge of the EoS parameters can help mitigate the uncertainties, the high mass end suffers from larger uncertainties, because its constraints rely solely on the GW data. Finally, we note that the larger number of events helps to reduce the uncertainties by accumulating more events, but it does not change the overall shape of the uncertainty distribution.

\subsubsection{What information is communicated about the EoS?}
In this section, we expand upon the discussion of Fig.~\ref{fig:b3} above, to illustrate how GW constraints inform the
EoS, mass-radius relation, and underlying EoS model parameters. 

Figure \ref{fig:deos} shows the relative uncertainties of pressure as a function of number density. Between
$\sim 1.3$-$2\,n_0$, the relative uncertainty in pressure is small.  This region's location is largely independent of
the mass distribution models because  this density range roughly corresponds to the central densities of NSs with masses
around $1.4\,\Msun$. As discussed above, these source masses provide the best relative constraints on TD,  due to having
masses large enough to have appreciable TD.
At higher densities, the uncertainties of pressure start to increase with density, as the tidal effects weaken for
massive NSs.
At sub-saturation densities including the inner crust region, however, the relative uncertainty in the pressure
increases markedly.  This large relative uncertainty reflects the fact that
the TD of low mass NSs are not very sensitive to the corresponding EoS: even $\sim 5\%$ uncertainties in TD can lead to
more than $30\%$ uncertainties in pressure at sub-saturation densities.

For comparison, Fig.~\ref{fig:deos} also includes the current EoS uncertainties at low densities from
Ref.\citep{2024A&A...687A..44D}, shown as the dark red shaded region.
We find that neither bimodal nor uniform models provide tighter constraints than the current knowledge from nuclear
physics and observations. The combination of these complementary information with GW constraints may significantly
improve the EoS constraints at sub-saturation densities.

Figure \ref{fig:radius_dradius} shows the 90\% CI of the radius posterior, expressed relalative uncertainty in the radius (left panel)
The radius uncertainties exhibit a flatter trend compared to those of the TD and pressure, and remaining at a level of $\sim 400$m across the whole mass ranges. Unlike the TD, radius of NS cannot be directly measured from the GW signals. Although a universal relation between radius and TD~\citep{2017PhR...681....1Y} exist, its intrinsic errors are non-negligible at the precision achieved by 3G detectors. Consequently, the uncertainties of TD cannot straightforwardly translate to radius uncertainties. On the other hand, the radius of $\sim 1.4\,\Msun$ NS is sensitive to the sub-saturation and inner-crust EoS, which are not well constrained by the GW data alone. This further limits the accuracy of radius measurements at this mass region. In the right panel of Fig.~\ref{fig:radius_dradius}, we find that the maximum mass of NS exhibits a relatively small uncertainty, consistent with the smaller pressure uncertainties at high density region shown in Fig.~\ref{fig:deos}. This is partly because the abundance of samples near maximum mass is relatively higher than that at low mass end for both bimodal and uniform distributions, as can be seen in Fig.~\ref{fig:m_dist}. Additionally, the intrinsic limitations of RMF EoS may impose some constraints at high densities. The RMF-based EoS generally allow less freedom at high density compared to those parameterization methods with more flexibility~\citep{2010PhRvD..82j3011L, 2018PhRvL.121p1101A, 2021ApJ...919...11H, 2022PhRvD.105f3031L, 2022ApJ...926..196H, 2024PhRvD.110h3030L, 2024arXiv241014674Y, 2024arXiv240715753V}.

Figure ~\ref{fig:eos_params} shows the one-dimensional marginal  posterior distributions for each of our EoS paraemters;
in an appendix, Fig.~\ref{fig:corner_plot} provides the conventional ``corner plot'' grid of two-dimensional marginal
posterior distributions.   Because prior
information plays an important role in constraining the EoS for the range of detection numbers investigated here,  a
larger number of events does not significantly improve the constraints on most EoS parameters, except for $E_{\rm sym}$ and
$M_N^\ast / M_N$.  Rather than measurement accuracies scaling as $1/\sqrt{N}$, as would be expected in a data-dominated
regime, the one-dimensional and two-dimensional marginal distributions reflect both GW observations and our strong prior
knowledge about the EoS.    Too, as discussed in Ref.~\citep{2025arXiv250811875Z},  degneracies exist among the EoS
parameters, such that correlated combinations can produce similar effects on TD and the zero-temperature
beta-equlibrated EOS.   


\subsection{The impact of mass distributions}

The mass distribution of NSs in BNS systems is not well known. However, it plays a crucial role in determining the final EoS constraints from GW observations. In this subsection, we compare the resulting constraints on the EoS, TD, pressure, and radius for different mass distribution models.

We observe in Fig.~\ref{fig:mchi_lamtilde} that the different mass distribution models (bimodal vs. uniform) lead to similar SNRs, which is consistent with the SNR distribution in Fig.~\ref{fig:snr_dist}. Indeed, the distance is the dominant factor affecting the SNR of GW signals, while the mass distribution has only a minor impact. Meanwhile, the TD uncertainties also exhibit similar results for both mass distribution models. Due to the trade-off between mass and TD discussed in the previous subsection, the events with the best TD measurements are concentrated in the intermediate chirp-mass region, which is accessible to both mass models. Although the uniform models provide more low-mass NSs, whose TDs are larger and can be measured more accurately, and the best 200 TD measurement events in this model extend to lower chirp mass region, the overall TD uncertainties remain comparable to those of the bimodal model.

\begin{figure}[t!]
  {\centering
  \includegraphics[width=0.5\textwidth]{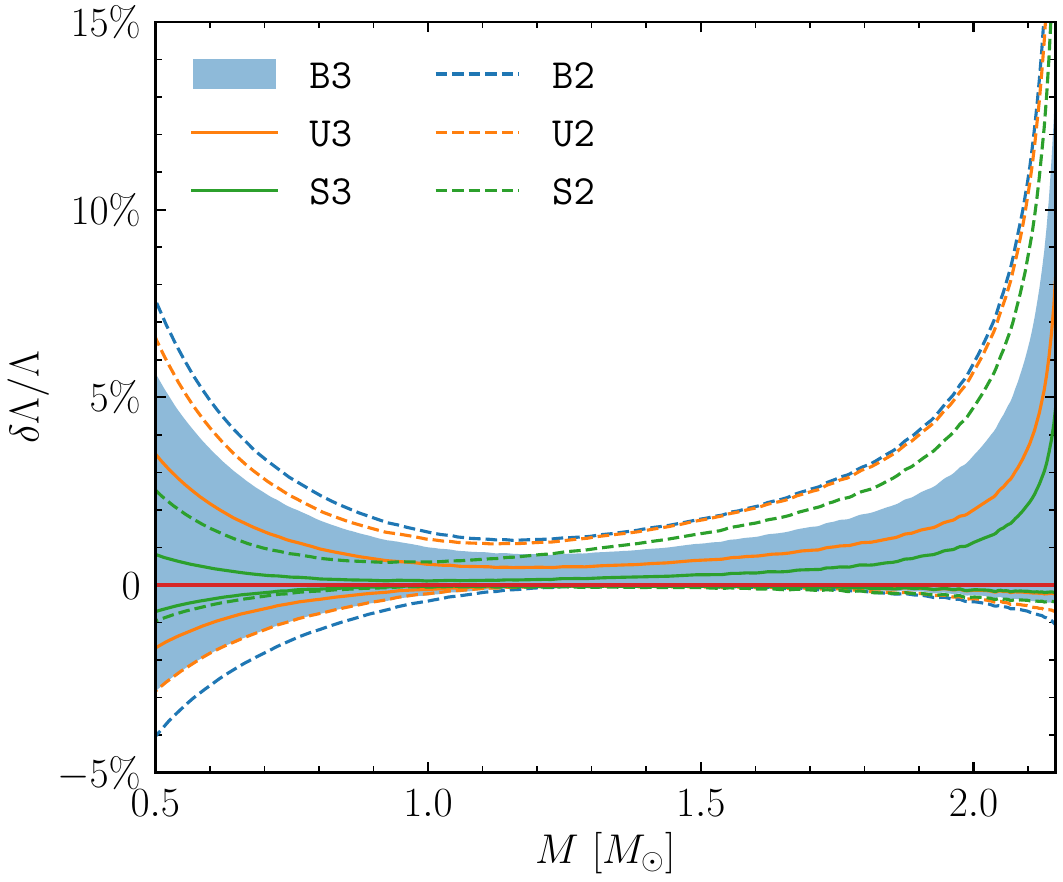}}
  \caption{The relative uncertainties in the tidal deformability, $\delta \Lambda/\Lambda$, as a function of NS mass for different models. The contour lines represent the 90\% credible intervals of the posterior distributions. The model labels are the same as those in Fig.~\ref{fig:mchi_lamtilde}.}
  \label{fig:dlambda}
\end{figure}

In Fig.~\ref{fig:dlambda}, the TD uncertainties as a function of NS mass for the uniform models shows profiles similar to those of the bimodal models, although the uniform models provide slightly smaller uncertainties across the entire mass range. The uniform models include more low-mass and massive NSs than the bimodal models, which helps reduce the TD uncertainties at both low and high mass ends. However, we find that the TDs at typical mass region, $1.1$-$1.5\,\Msun$, are still measured with higher precision in the uniform models than in the bimodal models, despite the latter containing more NSs in this mass range. Although not very clear, Fig.~\ref{fig:mchi_lamtilde} shows that events of the bimodal models at this mass region have relatively smaller TD uncertainties (darker colors) than the uniform models. Nevertheless, certain events in the bimodal models are effectively wasted for EoS constraints: they concentrate around $1.4\,\Msun$ and provide redundant information about the EoS, whereas events with low-mass and massive NSs in uniform models provide more informative leverage on the EoS and lead to smaller uncertainties in both the EoS and TDs.

Indeed, we see similar behavior in the EoS uncertainties presented in Fig.~\ref{fig:deos}. The uniform models yield smaller pressure uncertainties across all density regions than the bimodal models. Additionally, thanks to the abundance of low mass NSs in the uniform models, the constraints at sub-saturation densities and inner-crust region are improved, despite still being less informative than current knowledge from nuclear physics and observations. Finally, extremely tight constraints on the pressure at densities around $1.5\,n_0$ are observed for all models. This corresponds to NSs with masses around $1.4\,\Msun$, whose TD are best constrained due to their moderate TD contributions to the GW signal and their relatively larger SNRs.

As discussed previously, the radius is more sensitive to the low-density and inner-crust EoS than the TD. Therefore, no tight constraints on radius can be observed in Fig.~\ref{fig:radius_dradius}. The higher precision measurements of the TD around $1.4\,\Msun$ and of EoS around $1.5\,n_0$ do not translate into significantly tighter radius constraints in the corresponding mass region. Nevertheless, the uniform models still provide smaller radius uncertainties across the entire mass range than the bimodal models, due to their better constraints on the EoS at sub-saturation densities and inner-crust region. On the other hand, the uniform models also yield smaller uncertainties on the maximum NS mass because they include more massive NSs near maximum mass cutoff in the population.

Due to the higher precision of the EoS, we find that the constraints on the nuclear parameters are improved in the uniform models, as shown in Fig.~\ref{fig:eos_params}. However, these improvements are limited by parameter degeneracies. GW observations alone cannot achieve arbitrarily high precision in measuring the EoS parameters, even with larger number of events and wider mass distribution. We provide further discussion of these degeneracies in Appendix~\ref{sec:appendix2}.

In summary, we find that the mass distribution of NSs in BNS systems plays a crucial role in determining the final EoS, TD, and radius constraints from GW observations. A wider mass distribution, such as the uniform distribution considered in this work, includes more low-mass and massive NSs and therefore provide more information about the EoS. However, despite the abundance of $\sim 1.4\,\Msun$ NSs in the bimodal Gaussian distribution, they do not yield higher precision for NSs due to redundant information. Finally, we note that the impact of mass distribution of NSs can be more significant than that of the observed event population, which highlights the importance of understanding the mass distribution of NSs.

\subsection{Sub-solar mass neutron stars}

Although the sub-solar mass NSs (or black holes) are rare~\citep{2022NatAs...6.1444D} and none have been confirmed by GW observations so far, efforts to search their GW signals have never stopped~\citep{2023MNRAS.524.5984L, 2026arXiv260505444T}. The TDs of sub-solar mass NSs are much larger than those of typical NSs and are sensitive to the low-density and inner-crust EoS. Its observation can therefore provide unique information about the EoS at low densities that cannot be accessed by typical NSs. In this subsection, we will discuss the implications of sub-solar mass NSs for the final EoS constraints.

Because of the inclusion of sub-solar mass NSs, we observe the best 200 TD measurement events extend to a much lower chirp mass region in Fig.~\ref{fig:mchi_lamtilde}. For these sub-solar mass NSs, matter effects are significantly stronger and dominate the TD measurements. Hence, the trade-off between mass and TD discussed previously is not observed for the \texttt{S} models. Meanwhile, the \texttt{S} models also exhibit smaller TD uncertainties than the other two models, as indicated by the generally darker colors in this figure. Because we only show the best 200 TD measurement events, the sub-solar mass NSs contribute a considerable fraction of these events and also lower the uncertainties at other mass regions shown in the figure. Therefore, we see darker colors and smaller TD uncertainties in all mass region than other two models.

\begin{figure}[t!]
  \vspace{0.15cm}
  {\centering
  \includegraphics[width=0.5\textwidth]{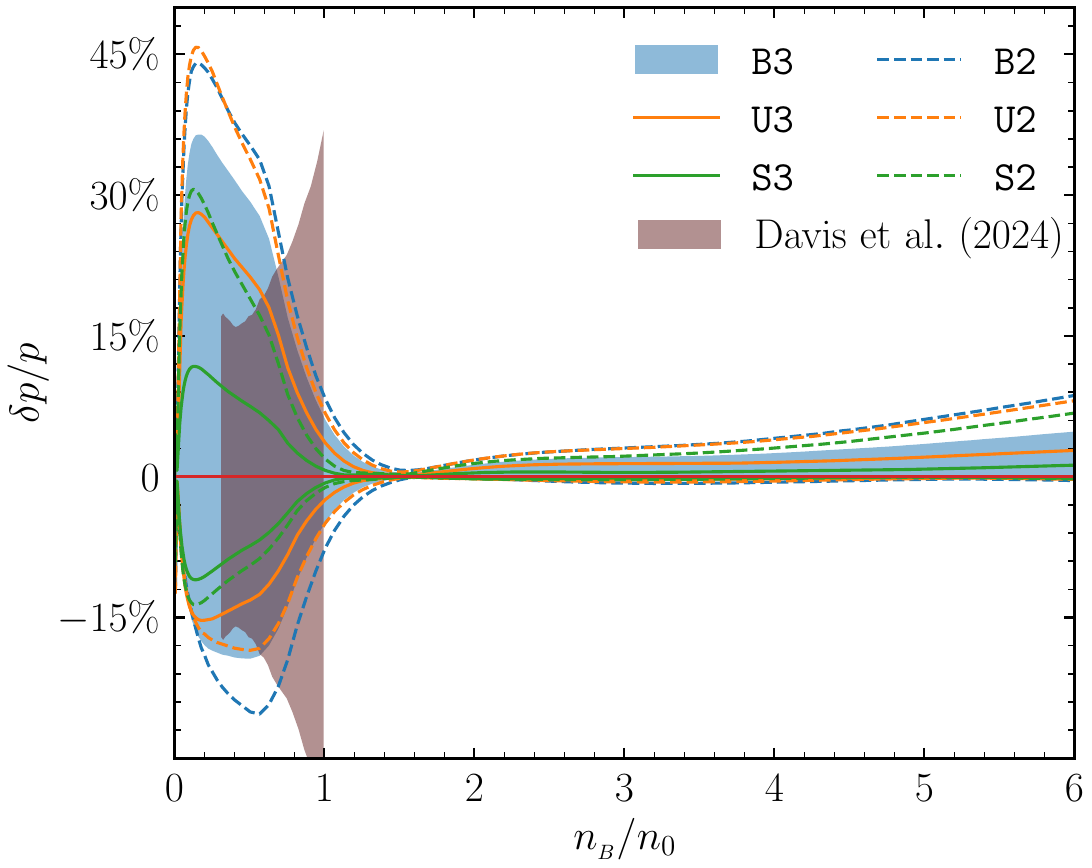}}
  \caption{The relative uncertainties in pressure as a function of number density. The labels are the same as in the previous figures. The results of sub-saturation EoS constraints from Ref.~\citep{2024A&A...687A..44D} are also displayed here for comparison, shown as the dark red shaded region.}
  \label{fig:deos}
\end{figure}

\begin{figure*}[t!]
  {\centering
  \includegraphics[width=0.49\textwidth]{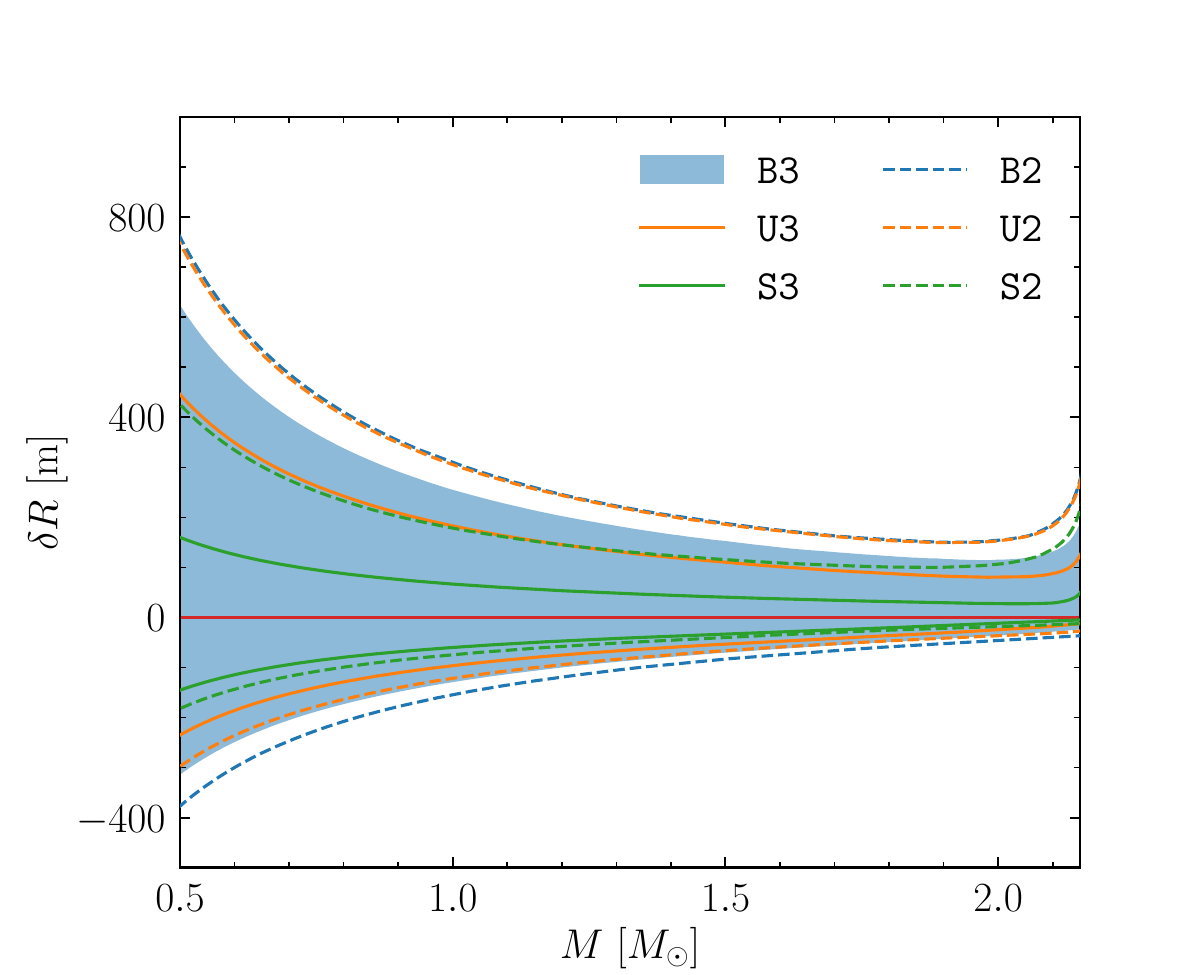}
  \includegraphics[width=0.49\textwidth]{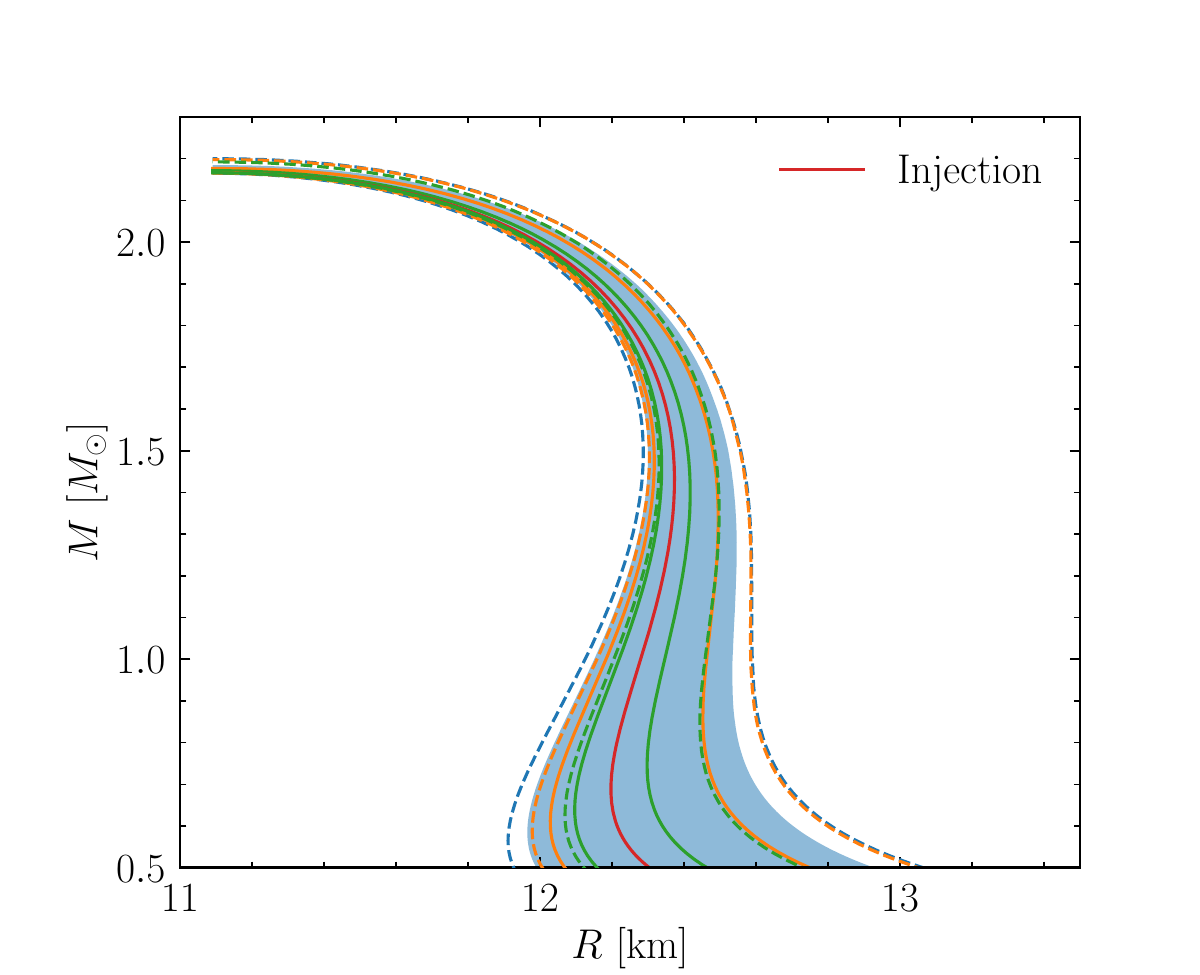}}
  \caption{The radius uncertainties as a function of NS mass (left) and the corresponding $90\%$ credible interval in mass-radius relation (right) for different models. The labels are the same as in the previous figures.}
  \label{fig:radius_dradius}
\end{figure*}

Compared to the other two models, the \texttt{S} models provide significantly smaller TD uncertainties across the entire mass range in Fig.~\ref{fig:dlambda}. In particular, the inclusion of sub-solar mass NSs helps reduce the TD uncertainties at the low mass end and can achieve a precision comparable to that in typical mass region. These low-mass or sub-solar mass NSs provide unique and important information about low-density EoS and nuclear parameters, and help reduce the overall TD uncertainties. However, although improvements are also seen at high mass end, the TD uncertainties for massive NSs above $2.0\,\Msun$ remain large due to the weakening of tidal effects.

The inclusion of sub-solar mass NSs significantly improves the EoS constraints at sub-saturation densities and in the inner-crust region, as shown in Fig.~\ref{fig:deos}. The \texttt{S3} model even yields higher precision than the current constraints from nuclear physics. This highlights the unique role of sub-solar mass NSs in probing the low-density EoS. Meanwhile, because of the tighter constraints at low-density, the overall EoS uncertainties are reduced across all density regions compared to the other two models.

The tighter low-density EoS constraints therefore lead to smaller radius uncertainties as shown in Fig.~\ref{fig:radius_dradius}. The \texttt{S3} model even yields radius uncertainties of approximately $100$m across the entire mass range, which is a significant improvement over the other two models. This demonstrates the crucial role of sub-solar mass NSs in constraining NS radii. However, given the rarity of sub-solar mass NSs (none have been confirmed through GW so far), constraining the low-density and inner-crust EoS through nuclear physics and radius measurements is are more efficient approaches than the GW observations. GW observations may provide complementary information if such sub-solar mass NSs are detected in the future, but they cannot be the primary means.

\begin{figure*}[t!]
  {\centering
  \includegraphics[width=0.33\textwidth]{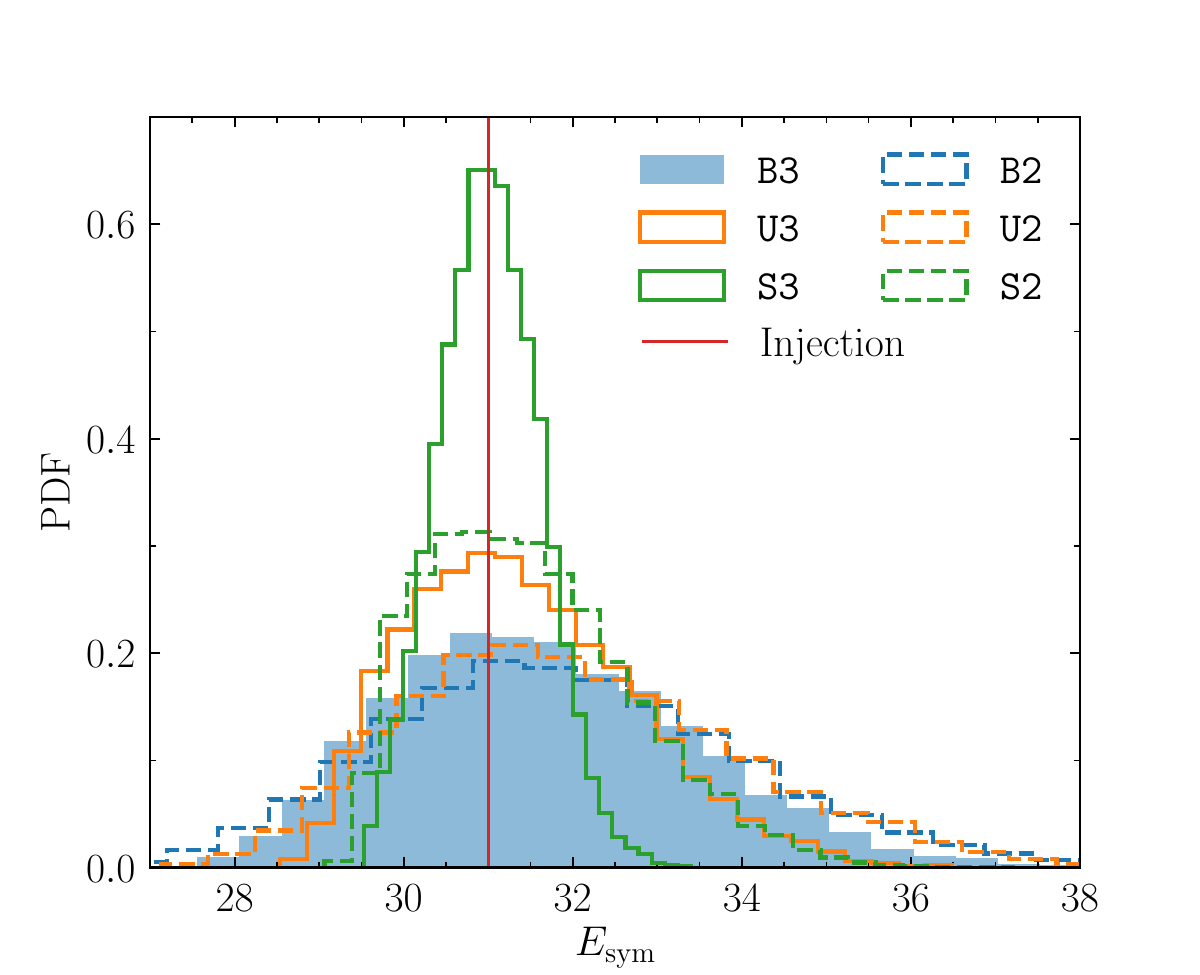}
  \includegraphics[width=0.33\textwidth]{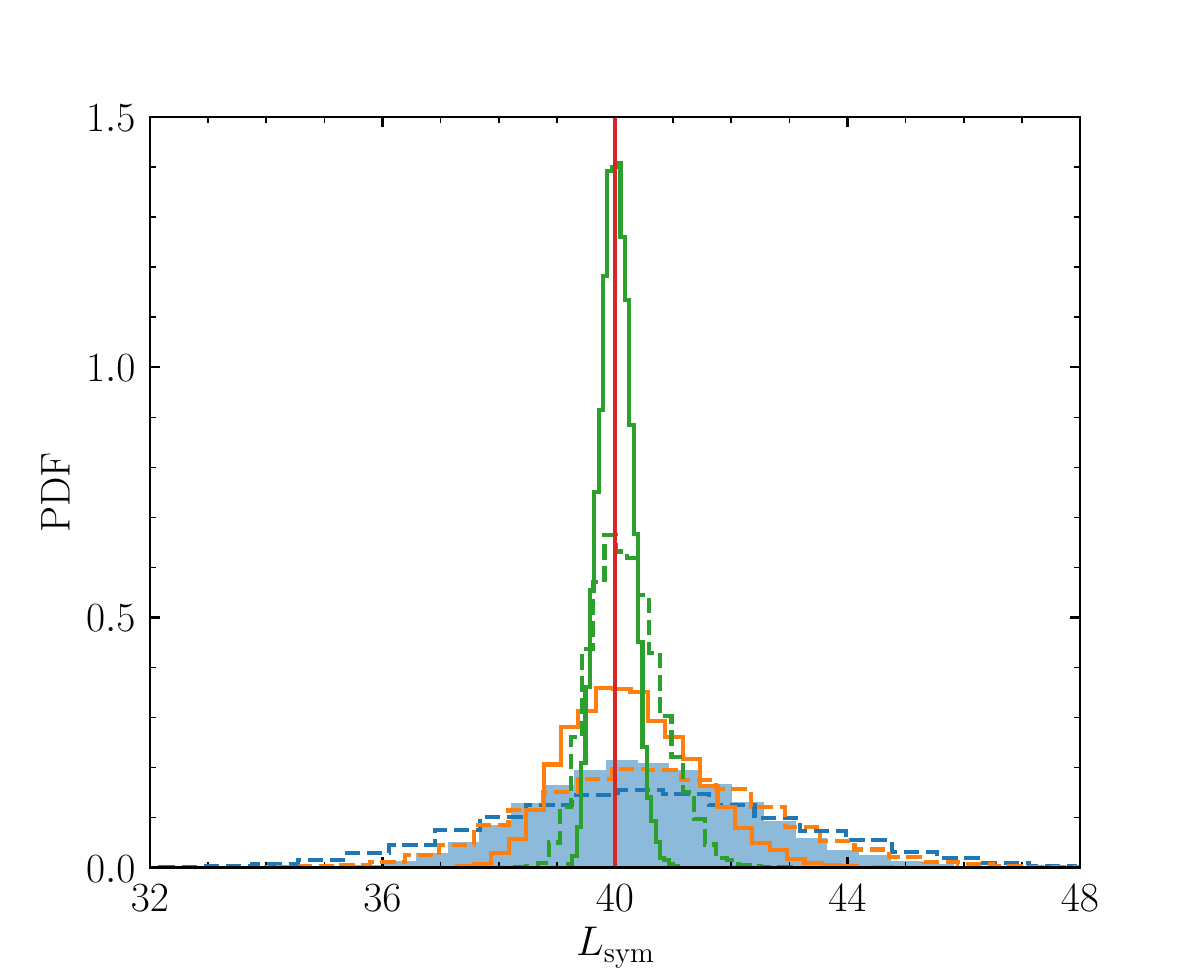}
  \includegraphics[width=0.33\textwidth]{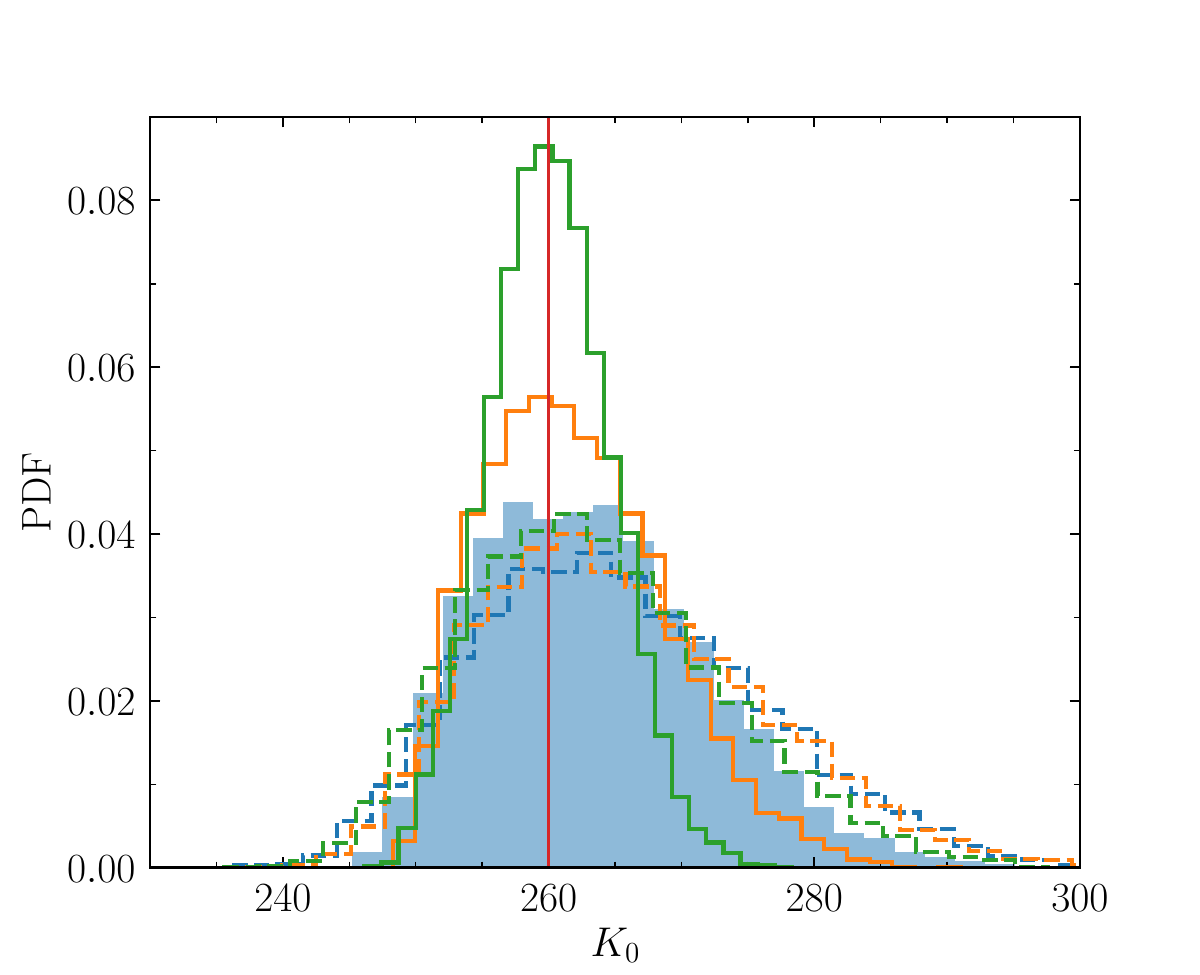}
  \includegraphics[width=0.33\textwidth]{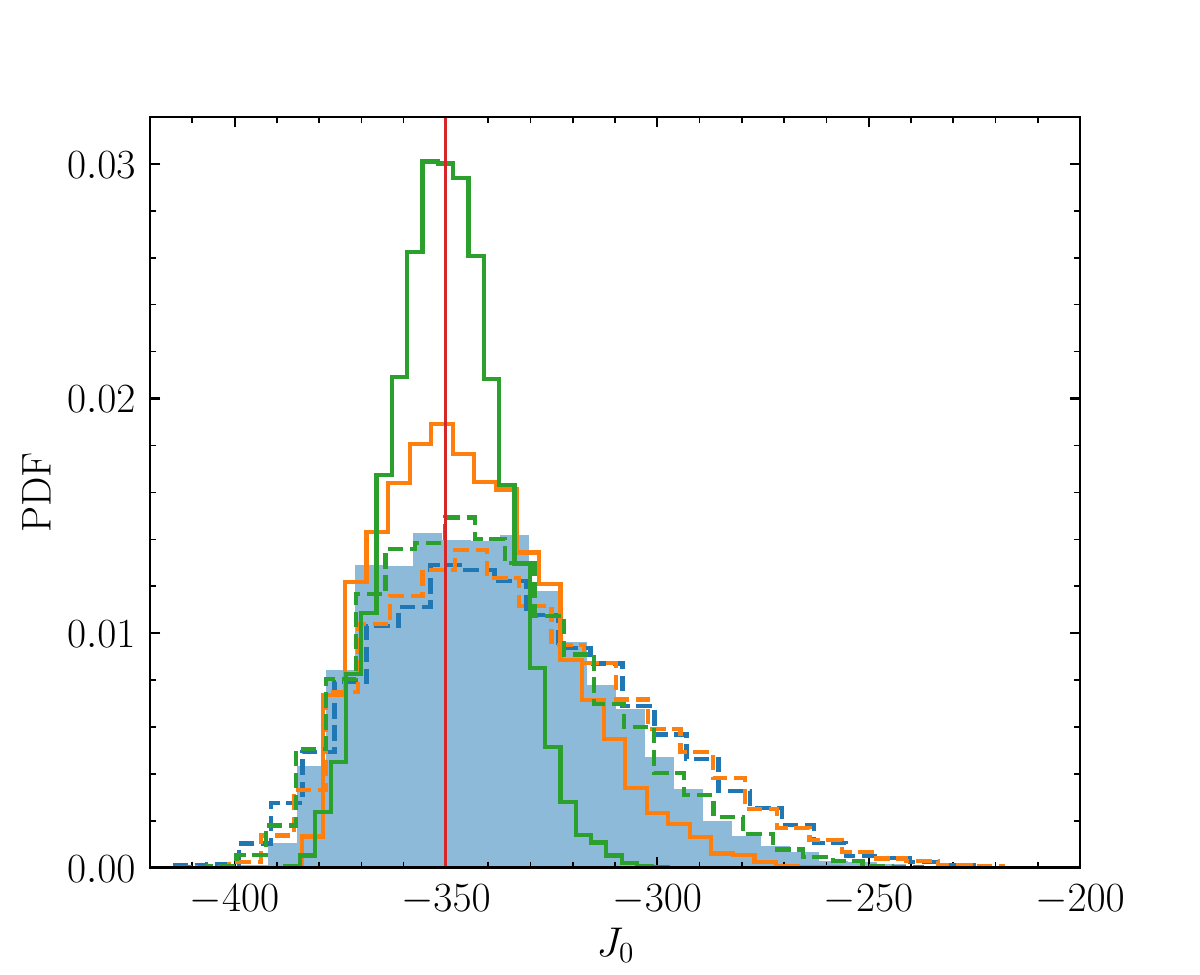}
  \includegraphics[width=0.33\textwidth]{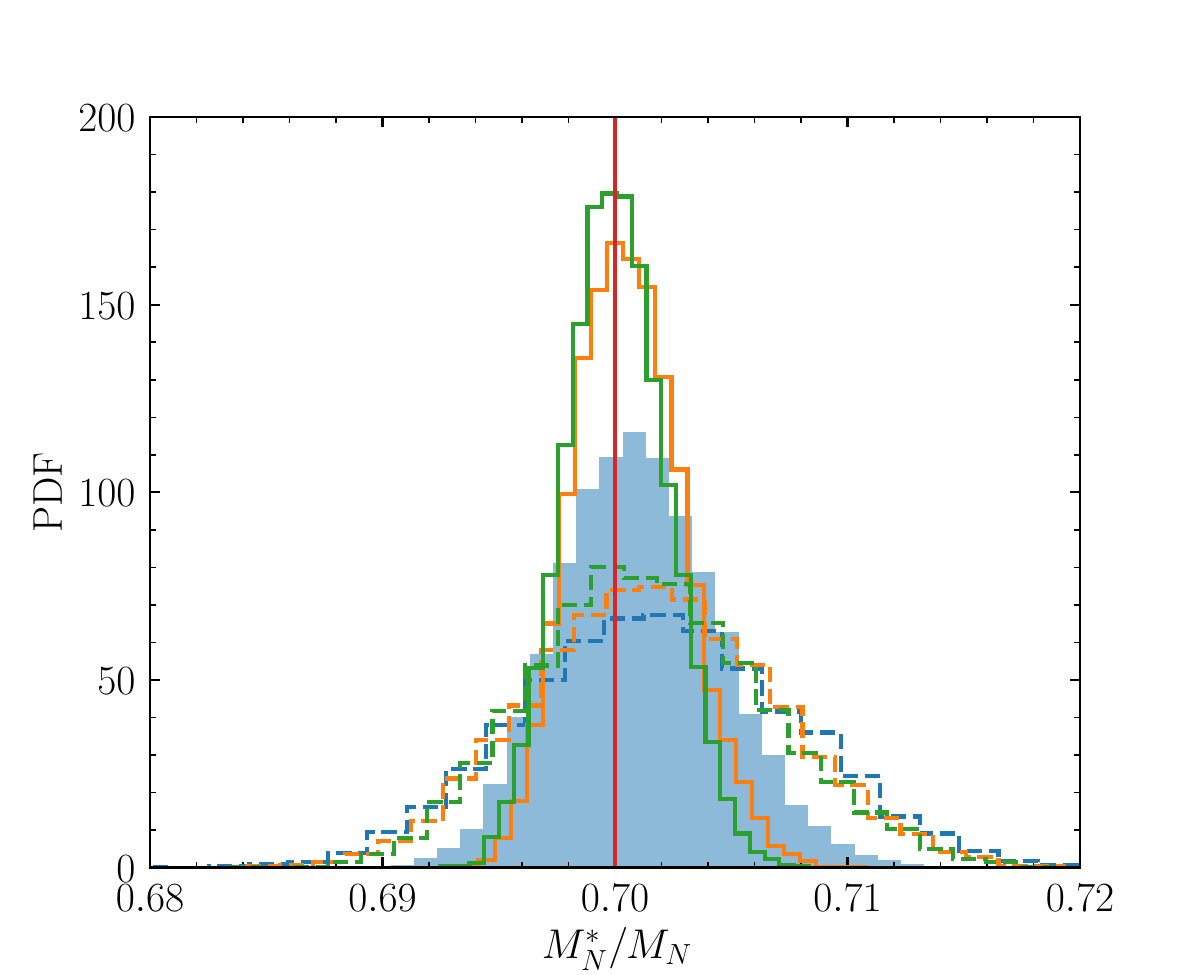}}
  \caption{The posterior distributions of the nuclear parameters for different models. The labels are the same as in the previous figures.}
  \label{fig:eos_params}
\end{figure*}

The \texttt{S3} model also provides the tightest constraints on the nuclear parameters, as displayed in Fig.~\ref{fig:eos_params}. In particular, for the posteriors of $L_{\rm sym}$, its uncertainties significant reduced due to its weaker degeneracies with other parameters~\citep{2025arXiv250811875Z}. However, the degeneracies among the EoS parameters still exist and prevent these nuclear parameters from being tightly constrained through GW observations alone. We discuss this in more detail in Appendix\ref{sec:appendix2}.

In general, the inclusion of sub-solar mass NSs can significantly improve the EoS, TD, radius, and nuclear parameter constraints from GW observations. These low-mass NSs provide unique information about the low-density and inner-crust EoS that cannot be accessed by typical NSs. However, given the rarity of sub-solar mass NSs, constraining the low-density and inner-crust EoS through nuclear physics and radius measurements is a more efficient approach than GW observations. This also implies the importance of multi-messenger observations in constraining the EoS.

\section{Conclusions}
\label{sec:conclusion}

We systematically investigated the impact of BNS merger event populations and NS mass distributions on constraining the EoS, TD and radius of NSs, as well as the nuclear parameters, using BNS inspiral GW signals with 3G detectors. Given the observed number of events and mass distribution models, we generated a set of simulated BNS merger GW signals with uniformly distributed sky locations and orientations and redshifts drawn from the cosmological merger distribution, and filtered out these with SNRs below 8. We then computed the Fisher information matrix for each event, held the non-EoS waveform parameters fixed at their injected values, and accumulated the FIM of EoS parameters over all events to obtain the final covariance of the EoS parameters. The posterior samples of EoS, TD, radius and nuclear parameters were then generated based on this covariance matrix.

We found that the merger events directly impact the observed event population. larger number of merger events leads to more observed events and higher SNRs of the loudest events, which indeed helps reduce the uncertainties of EoS, TD, radius and nuclear parameters through the accumulation of more data. However, the number of events do not change the overall shapes of the uncertainty distributions, but instead acts as a scaling factor. 

On the other hand, we note that the impact of mass distribution of NSs can be more significant than that of the event population. We compared the results from a realistic bimodal Gaussian mass distribution, a uniform mass distribution, and a uniform distribution that includes sub-solar mass NSs. Compared to the bimodal Gaussian distribution, the uniform distribution provides more low-mass and massive NSs, which help reduce the uncertainties of EoS, TD, radius and nuclear parameters across the entire mass and density ranges, these improvements can be more significant than those from larger number of events. This implies that the abundant $\sim 1.4,\Msun$ NSs in the bimodal Gaussian distribution provide partly redundant information. In particular, including sub-solar mass NSs can further improve the constraints, highlighting the importance of the low-density and inner-crust EoS. However, given the rarity of sub-solar mass NSs and current tight constraints from nuclear physics, constraining the low-density and inner-crust EoS through nuclear physics and radius measurements is a more efficient approach than GW observations. Better EoS constraints can therefore be expected by combining these complementary information from nuclear physics and radius measurements.

\begin{acknowledgements}
  ROS and ZZ acknowledge support from NSF AAG 2206321. ROS also acknowledges support from NSF PHY 2309172 and 2207920.
  This material is based upon work supported by NSF's LIGO Laboratory which is a major facility fully funded by the National Science Foundation.
  The authors are grateful for computational resources provided by the LIGO Laboratory and supported by National Science Foundation Grants PHY-0757058 and PHY-0823459.
\end{acknowledgements}

\bibliographystyle{apsrev4-1}
\bibliography{fisher2}

\appendix

\onecolumngrid
\section{The RHS of variational TOV and tidal equations}
\label{sec:appendix1}

We present the expressions for the RHS of the variational TOV and tidal equations here
\begin{gather}
\label{eq:derivatives}
  \mathcal{K}_{r^2, r^2} = -\frac{2z(1-2z)}{(4\pi r^2 p + z)^2},\ \ \ \mathcal{K}_{r^2, z} = \frac{2r^2(8\pi r^2 p +1)}{(4\pi r^2 p + z)^2},\ \ \ \mathcal{K}_{r^2, p} = \frac{8\pi r^4(1-2z)}{(4\pi r^2 p + z)^2},\ \ \ \mathcal{K}_{z, r^2} = -\frac{4\pi z(1-2z)(e+p)}{(4\pi r^2 p + z)^2}, \\
  \mathcal{K}_{z, z} = \frac{32\pi^2 r^4 ep - 16\pi r^2 pz - 2z^2 + 4\pi r^2 (e+p)}{(4\pi r^2 p + z)^2},\ \ \ \mathcal{K}_{z, p} = \frac{4\pi r^2(1-2z)(4\pi r^2 e-z)}{(4\pi r^2 p + z)^2},\ \ \ \mathcal{K}_{z, e} = -\frac{4\pi r^2(1-2z)}{4\pi r^2 p + z}, \\
  A_{H, r^2} = \frac{4\pi(1-2z)(e+3p)(z-4\pi r^2 p)}{(4\pi r^2 p + z)^3},\ \ \ A_{H, z} = -\frac{8\pi r^2(e+3p)(4\pi r^2 p - z + 1)}{(4\pi r^2 p + z)^3}, \\
  A_{H, p} = \frac{4\pi r^2(1-2z)(3z - 8\pi r^2 e - 12\pi r^2 p)}{(4\pi r^2 p + z)^3},\ \ \ A_{H, e} = \frac{4\pi r^2(1-2z)}{(4\pi r^2 p + z)^2},
\end{gather}
\begin{eqnarray}
\label{eq:derivatives1}
  B_{H, r^2} & = & \frac{(1-2z)(2z-8\pi r^2 p)}{(4\pi r^2 p + z)^3}\left(8\pi e+16\pi p+2\pi(p+e)\left(1+\kappa\right)\right) + \frac{48\pi p(1-2z)}{(4\pi r^2 p + z)^3}, \\
  B_{H, z} & = & -\frac{4r^2 (4\pi r^2 p + 1-z)}{(4\pi r^2 p + z)^3}\left(8\pi e+16\pi p+2\pi(p+e)\left(1+\kappa\right) - \frac{3}{r^2}\right), \\
  B_{H, p} & = & -\frac{16\pi r^4(1-2z)}{(4\pi r^2 p + z)^3}\left(8\pi e+16\pi p+2\pi(p+e)\left(1+\kappa\right) - \frac{3}{r^2}\right) + \frac{2r^2(1-2z)}{(4\pi r^2 p + z)^2} \left(16\pi+2\pi\left(1+\kappa\right) \right), \\
  B_{H, e} & = & \frac{2r^2(1-2z)}{(4\pi r^2 p + z)^2} \left(8\pi+2\pi\left(1+\kappa\right) \right), \\
  B_{H, \kappa} & = & \frac{4\pi r^2(1-2z)}{(4\pi r^2 p + z)^2} \left(e+p \right).
\end{eqnarray}

\section{The corner plot and degeneracies of EoS parameters}
\label{sec:appendix2}

\begin{figure*}[t!]
  {\centering
  \includegraphics[width=0.99\textwidth]{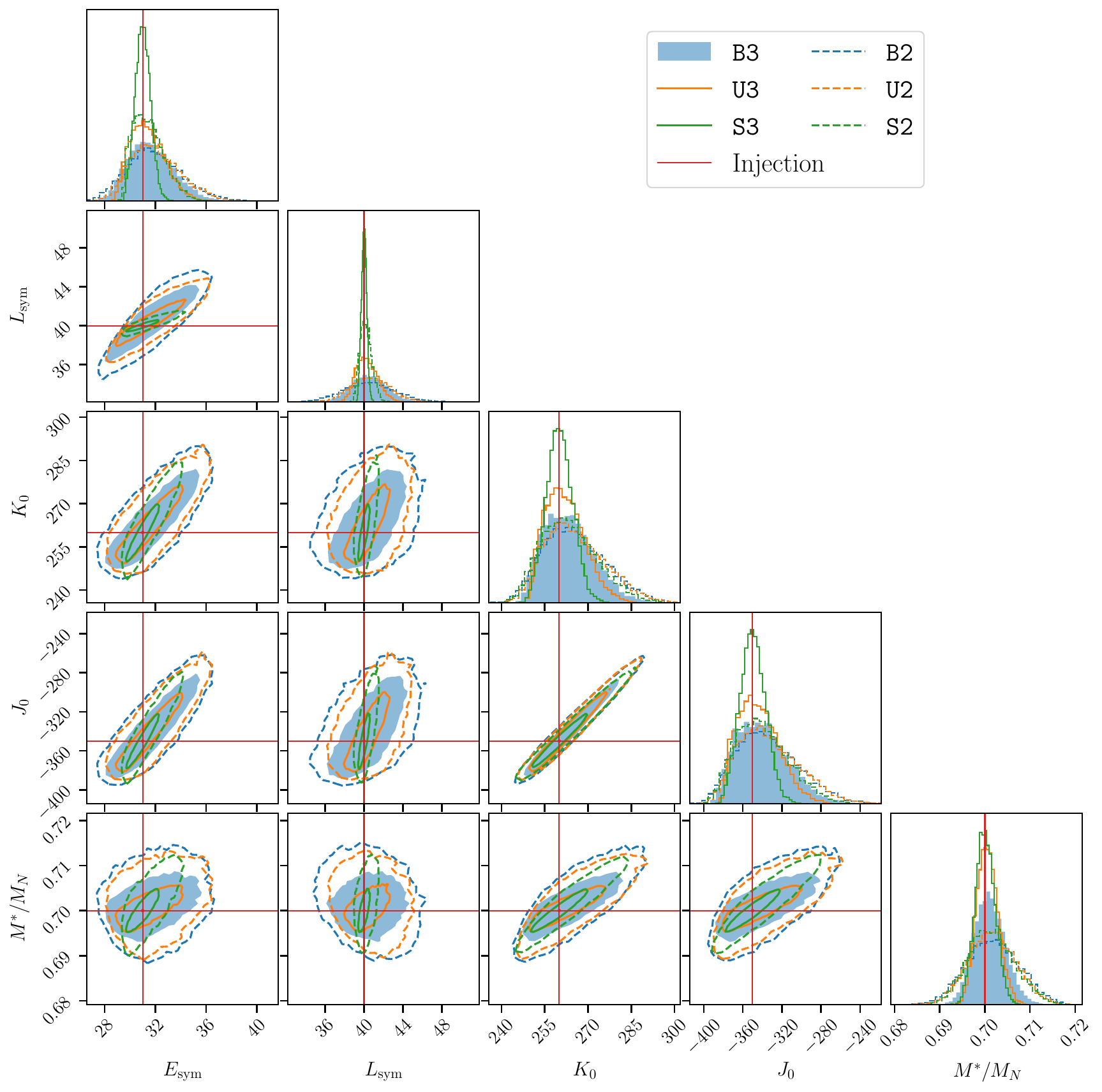}}
  \caption{The corner plot of nuclear parameters for different models. The labels are the same as in the previous figures.}
  \label{fig:corner_plot}
\end{figure*}

We present the corner plot of the nuclear parameters for different models in Fig.~\ref{fig:corner_plot}. We note that the contours of 90\% C.I. shrink as we increase the number of events or widen the mass distribution (from \texttt{B} to \texttt{U} to \texttt{S} models). However, the correlations always present between most of the parameters, implying the degeneracies among them. These degeneracies prevent the EoS parameters from being tightly constrained through GW observations alone, even with larger number of events and wider mass distribution. Additional information from nuclear physics experiments and multi-messenger observations are therefore necessary to break these degeneracies and achieve high-precision measurements of the EoS parameters.

\section{Estimation of maximum mass cutoff from mass distribution}
\label{sec:appendix3}

We have adopted a maximum mass cutoff in the mass distribution models, which can provide additional information about the EoS. This cutoff may be accurately determined through the NS mass distribution from GW data. We recall the likelihood for EoS hyperparameters 
after combining all single-event observations in Eq.~\eqref{eq:hierarchical}. By marginalizing over all parameters except the {\cal Y} and masses, we obtain
\begin{eqnarray}
  \label{eq:likelihood_pop}
  {\cal L}_{\rm net}({\cal Y}) & = & \prod_k \frac{\int d m_1 dm_2 \ell_k(m_1, m_2, {\cal Y})p(m_1|{\cal P}) p(m_2|{\cal P})}{\xi({\cal P})}\nonumber \\
  & = & \exp(-\alpha({\cal P}))\prod_k \ell_k^{\rm eos}({\cal Y}) \int d m_1 dm_2 \ell_k^{m}(m_1, m_2)p(m_1|{\cal P}) p(m_2|{\cal P}).
\end{eqnarray}
Where in the second line, we separate the EoS dependence because the EoS parameters are independent of the masses. The $m_1$ and $m_2$ are the masses of the two NSs in a BNS system, and ${\cal P}$ represents the hyperparameters of the mass distribution model. The likelihood $\ell_k$ are multivariate Gaussian distributions under the FIM approximation. The factor of selection effects $\xi({\cal P})$ is rewritten in the second line as $\exp(\alpha({\cal P}))$, and $\alpha({\cal P})$ represent the number of events to be detected given the mass distribution model. We recall in Fig.~\ref{fig:snr_dist} that the detected events are not sensitive to the mass distribution model, not even for the maximum mass cutoff. Therefore, we neglect the dependence of $\alpha({\cal P})$ on the $m_{\rm max}$ and treat it as a constant. The integral in the second can be evaluated after substituting the mass distribution model in Eq.~(\ref{eq:mass_dist}), and we show the example of the bimodal Gaussian model here. The likelihood can be written as
\begin{eqnarray}
  \label{eq:likelihood_pop2}
  {\cal L}_{\rm net}({\cal Y}) & \propto & \prod_k \ell_k^{\rm eos}({\cal Y}) \sum_{i,j=1,2} \int_{-\infty}^{m_{\rm max}} dm_1 dm_2 \ell_k^{m}(m_1, m_2) \left[r_{ij} \exp(-\frac{1}{2} \vec{m}-\mu_{ij})^T \sigma_{ij}^{-2} (\vec{m}-\mu_{ij})\right] \nonumber \\
  & = & \prod_k \ell_k^{\rm eos}({\cal Y}) \sum_{i,j=1,2} \int_{-\infty}^{m_{\rm max}} dm_1 dm_2 r_{ij} \exp(-K/2)\exp\left[-\frac{1}{2} (\vec{m}-\mu_{ij}^c)^T (\sigma_{ij}^{c})^{-2} (\vec{m}-\mu_{ij}^c)\right] \nonumber \\
  & = & \prod_k \exp\left[-\frac{1}{2} ({\cal Y}-\mu_{\rm eos})^T F^{\rm eos}_k ({\cal Y}-\mu_{\rm eos}) + \log Y(m_{\rm max})\right].
\end{eqnarray}
Where $\vec{m} = (m_1, m_2)^T$, $\mu_{ij}=(\mu_i, \mu_j)^T$, $\sigma_{ij} = \text{diag}(\sigma_i, \sigma_j)$, $r_{ij} = r_i r_j / (\Phi_i \Phi_j)$, and $K$ is an event-dependent constant. The $\mu_{ij}^c$ and $\sigma_{ij}^{c}$ are the mean and covariance of the combined Gaussian distribution after multiplying the likelihood and mass distribution model. The truncated integral can be calculated numerically and is denoted by $Y$ in third line. $F^{\rm eos}_k$ is the FIM for EoS parameters of the $k$-th event, and $\mu_{\rm eos}$ is the mean of EoS parameters.

We note that the contribution of the mass distribution to the EoS (or maximum mass) constraints arises from the truncated integral $Y(m_{\rm max})$. It is a function of the maximum mass $m_{\rm max}$ and therefore depends on the EoS parameters. We can further apply the linearized-signal approximation of FIM to $Y$ and write its contribution to final fisher matrix as
\begin{eqnarray}
  \label{eq:likelihood_pop2}
  \log ({\cal L}_{\rm net}({\cal Y})) \propto ({\cal Y}-\mu_{\rm eos})^T \sum_k\left(F^{\rm eos}_k + \frac{\partial^2 \log Y}{\partial {\cal Y}^2}\right) ({\cal Y}-\mu_{\rm eos}).
\end{eqnarray}
However, the computation of the second derivative of $Y$ with respect to EoS parameters is non-trivial, we simplify this procedure by calculating the Fisher matrix for the maximum mass $m_{\rm max}$ itself, $F^m=\sum_k \partial^2 Y/\partial m_{\rm max}^2$, and estimating its standard deviation.

Another issue is the distinguishability of NSs from BHs. The GW signals cannot distinguish between NSs and BHs if the TD measurements are not accurate enough. An observed compact object with a mass near the maximum mass cutoff could be either a massive NS or a low-mass BH. Therefore, in computation of $F^m$, we exclude the contributions from the compact objects with large TD uncertainties that cannot be distinguished from BHs. We adopt a threshold of $3\sigma$, \ie $\Lambda_{\rm mean} - 3\delta \Lambda < 0$, and include only events with a $3\sigma$ certification of a non-zero TD in the Fisher-matrix calculation for $m_{\rm max}$. 

In Table~\ref{tab:delta_mmax_data}, we present the fraction of distinguishable BNS events and the standard deviation of the maximum mass cutoff, $\delta m_{\rm max}$, estimated from the GW observations for all population models. We find that only $\sim 20\%$ of the events are distinguishable. On the other hand, the uncertainty in the maximum-mass cutoff is $\sim 0.1$. Massive NSs near the maximum-mass cutoff have smaller TDs and are therefore more difficult to distinguish from BHs. However, these massive NSs also carry more information about the maximum mass and are thus more important for constraining $m_{\rm max}$. Consequently, The constraints on $m_{\rm max}$ strongly depend on a small number of ``lucky" events containing massive NSs with high SNRs.

\begin{figure*}[t!]
  {\centering
  \includegraphics[width=0.6\textwidth]{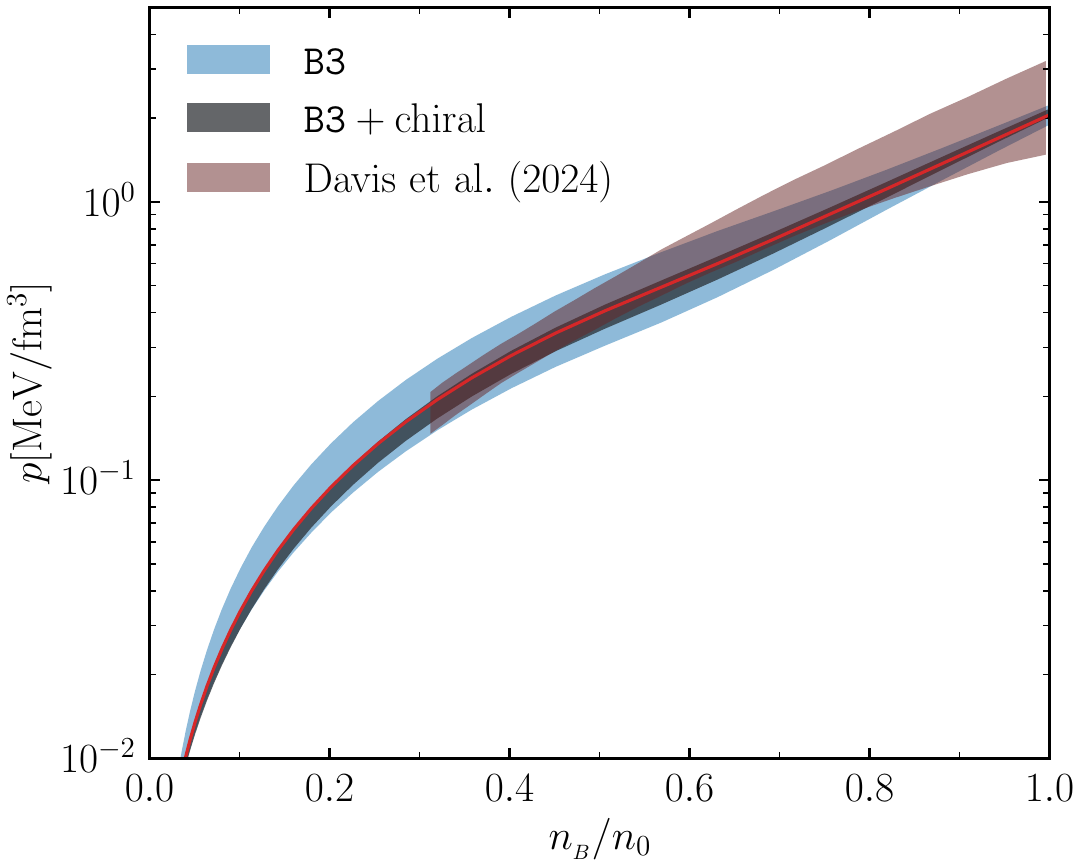}}
  \caption{The $90\%$ C.I. of the EoS posterior distribution.}
  \label{fig:chiral_post}
\end{figure*}

\begin{table}[t]
\begin{center}
\begin{tabular}{ccccccc}
\hline
Model &  \texttt{B2} & \texttt{B3} & \texttt{U2} & \texttt{U3} & \texttt{S2} & \texttt{S3} \\
\hline
$\delta m_{\rm max}\ [M_\odot]$ & $0.0548$ & $0.0161$ & $0.313$ & $0.143$ & $1.280$ & $0.0913$ \\
Distinguishable & \multirow{2}{*}{$21.0\%$} & \multirow{2}{*}{$21.0\%$} & \multirow{2}{*}{$17.5\%$} & \multirow{2}{*}{$17.5\%$} & \multirow{2}{*}{$14.7\%$} & \multirow{2}{*}{$14.8\%$} \\
events fraction & & & & & & \\
\hline
\end{tabular}
\end{center}
\caption{The standard deviation of the maximum mass cutoff, $\delta m_{\rm max}$, estimated from the GW observations. The results are obtained by calculating the Fisher matrix for $m_{\rm max}$ and taking the square root of its inverse.
}
\label{tab:delta_mmax_data}
\end{table}

\section{The constraints from chiral effective field theory}
\label{sec:appendix4}

The sub-saturation and inner-crust EoS can be further constrained by nuclear physics such as the Chiral effective field theory ($\chi$ EFT)~\citep{2016PhRvC..93e4314D}. However, the $\chi$ EFT constraints for pure neutron matter are typically provided as a distribution of pressure at each density point, and cannot be directly incorporated into our FIM framework. A fully MCMC inference would be required to accurately and consistently incorporate these constraints. In order to estimate its effects on the posterior EoS, we include the $\chi$ EFT information in a simplified way during the sampling process: we calculate the mean and standard deviation of the $\chi$ EFT pressure at each density point, and discard these EoS samples whose average pressure deviation from the $\chi$ EFT mean exceeds twice the $\chi$ EFT deviation.

The $90\%$ credible interval of posterior EoS with and without the $\chi$ EFT constraints are shown in Fig.~\ref{fig:chiral_post}. We note that the sub-saturation and inner-crust EoS constraints are significantly improved when the $\chi$ EFT information is included, and the posterior EoS with $\chi$ EFT constraints becomes consistent with the results from Ref.~\citep{2024A&A...687A..44D}. This highlights the importance of nuclear physics in constraining the low-density and inner-crust EoS.

\section{Prospects with FRIB-era laboratory symmetry-energy constraints}
\label{sec:appendix5}

The prior FIM adopted in Sec.~\ref{sec:methodology} is deliberately broad: it regularizes the matrix inversion without
materially constraining the EoS parameters, with one exception.  Because the GW data only weakly constrain
$E_{\rm sym}$ (cf.~Appendix~\ref{sec:appendix2}), even the broad $3\,{\rm MeV}$ prior width contributes appreciably to
the $E_{\rm sym}$ posterior: for the fiducial \texttt{B3} model, removing the prior entirely inflates
$\sigma(E_{\rm sym})$ from $2.22$ to $3.65\,{\rm MeV}$.  Precisely this parameter sector is the target
of current and planned rare-isotope laboratory programs, notably at the Facility for Rare Isotope Beams
(FRIB)~\citep{2024PrPNP.13404080S, 2024arXiv241006144B, 2026arXiv260629588L}.  To gauge how such measurements would
interact with the GW constraints forecast here, we replace the diagonal $(E_{\rm sym}, L_{\rm sym})$ entries of the
prior FIM with correlated $2\times 2$ inverse-covariance blocks representing two scenarios.  We emphasize that no
published projection commits FRIB to a specific precision on a specific date; the widths below are labeled scenario
assumptions of ours, chosen to be consistent with the program descriptions in
Refs.~\citep{2024PrPNP.13404080S, 2024arXiv241006144B, 2026arXiv260629588L}, not literature values.

In the first scenario, denoted \texttt{FRIB-2028}, the subsaturation program (masses, isobaric-analog states, dipole
response) together with first heavy-ion-collision transport results yields $\sigma(E_{\rm sym}) = 2\,{\rm MeV}$ and
$\sigma(L_{\rm sym}) = 25\,{\rm MeV}$ with correlation coefficient $r = 0.7$, reflecting the well-known positive
$E_{\rm sym}$--$L_{\rm sym}$ correlation of subsaturation observables.  In the second scenario, denoted
\texttt{FRIB400}, a suprasaturation measurement campaign (representative of the proposed FRIB400 upgrade era, early
2030s) additionally determines the symmetry energy at twice saturation density, $E_{\rm sym}(2n_0)$, to
$\pm 10$--$15\,{\rm MeV}$.  We map this constraint into the saturation parameters through the RMF model itself:
linearizing $E_{\rm sym}(2n_0)$ about the injected EoS, holding $(K_0, J_0, M^*_N)$ fixed, gives
$\partial E_{\rm sym}(2n_0)/\partial E_{\rm sym} = 0.81$ and
$\partial E_{\rm sym}(2n_0)/\partial L_{\rm sym} = 0.14\,$, and the corresponding rank-one block
$\vec{g}\vec{g}^T/\sigma^2$ is added to the \texttt{FRIB-2028} prior FIM.  Note that these derivatives differ
substantially from the density-expansion values $(1, 1/3)$ because, at fixed $(K_0, J_0, M^*_N)$, the RMF model ties
the higher symmetry-energy coefficients (e.g.\ $K_{\rm sym}$) to $(E_{\rm sym}, L_{\rm sym})$.

Table~\ref{tab:frib_scenarios} lists the resulting forecast uncertainties for the \texttt{B2} and \texttt{B3} models.
Three features stand out.  First, the \texttt{FRIB-2028} block tightens $\sigma(E_{\rm sym})$ by
$38\%$ ($45\%$) for \texttt{B3} (\texttt{B2}) relative to the current broad prior --- and,
because the $E_{\rm sym}$ direction participates in the leading parameter degeneracies discussed in
Appendix~\ref{sec:appendix2}, this single laboratory input propagates to \emph{all five} saturation parameters, whose
uncertainties shrink by $\sim 25$--$35\%$.  The correlated off-diagonal term matters: with $r = 0$ the
$E_{\rm sym}$ improvement for \texttt{B3} drops from $38\%$ to $23\%$, since the correlation
transfers part of the precise GW $L_{\rm sym}$ measurement onto $E_{\rm sym}$.  Second, the GW data dominate the
$L_{\rm sym}$ constraint in all scenarios: even the assumed laboratory width $\sigma(L_{\rm sym}) = 25\,{\rm MeV}$ is
an order of magnitude broader than the GW-only constraint, so the direct laboratory information acts almost entirely
through $E_{\rm sym}$.  Third, the \texttt{FRIB400} suprasaturation constraint adds little further information
\emph{within the RMF parameter space} ($\lesssim 1\%$): once $(E_{\rm sym}, L_{\rm sym})$ are constrained, the model
itself predicts $E_{\rm sym}(2n_0)$ far more precisely than the assumed $\pm 10\,{\rm MeV}$ measurement.  The value of
such a measurement is therefore not parameter estimation within the model but an experimental test of the model's
density extrapolation --- exactly the regime where mean-field extrapolations are least trustworthy --- which a
within-model Fisher analysis cannot quantify.  None of the conclusions of the main text are altered by these
scenarios: the parameter degeneracies persist, and the laboratory information acts on the directions GW observations
leave unconstrained, underscoring the complementarity of the two channels.  For the sub-solar-mass models
(\texttt{S3} in particular), the GW data already dominate every parameter and the scenario priors change the
forecasts by only a few percent.

\begin{table*}[t]
\begin{center}
\begin{tabular}{lccccc}
\hline
 & $\sigma(E_{\rm sym})$ & $\sigma(L_{\rm sym})$ & $\sigma(K_0)$ & $\sigma(J_0)$ & $\sigma(M^*_N/M_N)$ \\
 & [MeV] & [MeV] & [MeV] & [MeV] & \\
\hline
\multicolumn{6}{c}{\texttt{B2}}\\
\hline
GW only (no prior) & 7.22 & 6.45 & 34.6 & 107 & 0.0107 \\
broad prior (current draft) & 2.61 & 2.76 & 14.1 & 42 & 0.0068 \\
\texttt{FRIB-2028} & 1.45 & 1.99 & 10.0 & 27 & 0.0063 \\
\texttt{FRIB400} ($\pm 15\,$MeV) & 1.44 & 1.99 & 10.0 & 27 & 0.0063 \\
\texttt{FRIB400} ($\pm 10\,$MeV) & 1.44 & 1.98 & 10.0 & 27 & 0.0063 \\
\hline
\multicolumn{6}{c}{\texttt{B3}}\\
\hline
GW only (no prior) & 3.65 & 3.28 & 17.4 & 54 & 0.0053 \\
broad prior (current draft) & 2.22 & 2.10 & 10.9 & 33 & 0.0040 \\
\texttt{FRIB-2028} & 1.37 & 1.43 & 7.4 & 22 & 0.0035 \\
\texttt{FRIB400} ($\pm 15\,$MeV) & 1.36 & 1.43 & 7.4 & 22 & 0.0035 \\
\texttt{FRIB400} ($\pm 10\,$MeV) & 1.36 & 1.42 & 7.3 & 22 & 0.0035 \\
\hline
\end{tabular}
\end{center}
\caption{Forecast standard deviations of the nuclear saturation parameters for the \texttt{B2} and \texttt{B3} models
under the labeled FRIB scenario priors, compared with the GW-only constraint (no prior) and the broad regularization
prior used in the main text.  The scenario widths are assumptions of ours (see text), not published projections.  All
values are Gaussian (Fisher) standard deviations, before the physicality cuts applied to the posterior samples shown
in the figures.}
\label{tab:frib_scenarios}
\end{table*}

\end{document}